\documentclass[10pt]{article}
\usepackage{setspace}
\usepackage{graphicx}
\usepackage{caption}
\usepackage{subcaption}
\usepackage{graphics}
\usepackage{amscd}
\usepackage{array}
\usepackage{amsfonts}
\usepackage{fancyhdr}
\usepackage{oldlfont}
\usepackage{latexsym,amssymb,amsmath}
\usepackage{amsthm}

\newtheoremstyle{theoremd}
  {}
  {}
  {\em}
  {}
  {\bf}
  {.}
  { }
  {\thmname{#1}\;\thmnumber{#2}\thmnote{\textnormal{ (#3)}}}
\theoremstyle{theoremd}

\newtheorem{theorem}{Theorem}

\newtheorem{corollary}{Corollary}

\newtheorem{lemma}{Lemma}

\newtheorem{proposition}{Proposition}
\newtheorem{remark}{Remark}

\def \Prod{\displaystyle\prod}

\def \be{\begin{eqnarray}}
\def \ee{\end{eqnarray}}
\def \b*{\begin{eqnarray*}}
\def \e*{\end{eqnarray*}}

\def \RR{{\mathbb R}}
\def \ZZ{{\mathbb Z}}

\def \Bc{{{\mathcal B}}}
\def \Rc{{{\mathcal R}}}
\def \Nc{{{\mathcal N}}}

\def \xx{{\mathbf{x}}}
\def \yy{{\mathbf{y}}}

\def \vv{{\mathbf{v}}}

\def \[{[\,\!\![}
\def \]{]\,\!\!]}

\def \1{{\bf 1}}

\def\Bc{{\cal B}}

\def\Ic{{\cal I}}

\def\Nc{{\cal N}}

\def\Rc{{\cal R}}

\def\mds{\medskip}

\def\eps{\varepsilon}

\makeatletter
\renewcommand*{\@fnsymbol}[1]{\ensuremath{\ifcase#1\or *\or **\or \dagger\or \ddagger\or
   \mathsection\or \mathparagraph\or \|\or \dagger\dagger
   \or \ddagger\ddagger \else\@ctrerr\fi}}
\makeatother

\begin{document}

\title{{\Large \bf ON THE STATIONARITY OF DYNAMIC CONDITIONAL CORRELATION MODELS}}
\medskip
\author{{\sc Jean-David Fermanian} \\ {\it Crest-Ensae}\thanks{3 avenue Pierre Larousse, 92245 Malakoff cedex, France; jean-david.fermanian@ensae.fr}
 \\ {\sc Hassan Malongo} \\ {\it Amundi \& Univ. Paris
Dauphine}\thanks{Univ. Paris-Dauphine, Place du Mar\'echal de Lattre
de Tassigny, 75116 Paris, France; hassan.malongo@amundi.com}}

\maketitle

\begin{abstract}
We provide conditions for the existence and the uniqueness of strictly stationary solutions of the usual Dynamic Conditional Correlation
GARCH models (DCC-GARCH). The proof is based on Tweedie's (1988) criteria, after having rewritten DCC-GARCH models as nonlinear
Markov chains. We also study the existence of their moments and discuss the tightness of our sufficient conditions.
\medskip
\noindent

\medskip
\noindent

\end{abstract}

\pagebreak

\section{Introduction}

\subsection{The problem}

In multivariate extensions of GARCH models, modelers are faced with the problem of correlations (between asset returns, in most applications).
The simplest idea is to assume that these correlations are constant in time, and constitute only an additional matrix of parameters.
This has provided the class of Constant Conditional Correlations models (CCC), first introduced by Bollerslev (1990). Since CCC models can be seen as the components of first-order  Markov processes, once such models are rewritten in an extended vector space, it is relatively easy to prove the existence of strictly stationary and explicit solutions, even if the latter are analytically complex: see classical textbooks,
for instance Francq and Zako\"{\i}an (2010).

\mds

It rapidly became apparent that the assumption of constant correlations is too strong. It does not correspond to economic intuition or many
empirical features: see the recent paper by Otranto and Bauwens (2013) and the numerous references therein, for instance. Therefore,
Engle (2002) proposed extending CCC specifications by adding particular dynamics on the (conditional) correlation matrices of returns, denoted here by $(R_t)$. To ensure modelers are dealing with true correlation matrices, he introduced a non-linear transform: there exists a sequence of
variance-covariance matrices $(Q_t)$ such that $R_t =diag(Q_t)^{-1/2} Q_t diag(Q_t)^{-1/2}$, and $(Q_t)$-dynamics are specified instead of $(R_t)$-dynamics directly, contrary to other authors (Tse and Tsui, 2002 or Pelletier, 2006, for instance). This non-linear transform ensures that $R_t$ is always a correlation matrix, i.e. positive semidefinite with ones on its main diagonal. Nonetheless, it considerably complicates the work of stating DCC model stationarity conditions. Indeed, analytically tractable solutions to such processes no longer exist. This explains why the existence and uniqueness of DCC model stationarity solutions have not yet been established in the literature, nor have the finiteness of their
moments. Particularly, this implies that theoretically sound statistical inference procedures do not yet exist, as noted in Caporin and McAleer (2013).

\mds

Despite their theoretical shortcomings, DCC models have been used intensively among academics and practitioners. Besides numerous applied works, several extensions of the baseline DCC representation have been proposed in the literature: inclusion of asymmetries (Cappiello, Engle, and Sheppard, 2006), volatility thresholds (Kasch and Caporin, 2013), macro-variables (Otranto and Bauwens, 2013), univariate switching regime probabilities (Pelletier, 2006, Billio and Caporin, 2005, Fermanian and Malongo, 2013), among others. Other authors have revisited the DCC parameterization itself: Billio, Caporin, and Gobbo (2006), Franses and Hafner (2009), etc. Therefore, there is an urgent need for new theoretical
results concerning the seminal DCC model itself.

\mds

Usually in econometrics, proving the existence of stationary solutions is the first step towards developing a full asymptotic theory (consistency/asymptotic normality of QML estimates typically, as in Comte and Lieberman (2003) in the case of multivariate GARCH models), because laws of large numbers and some CLTs are easily obtained in this case. In the GARCH literature, this essential task has been fulfilled by Bougerol and Picard (1992) for univariate GARCH models, by Ling and McAleer (2003) for multivariate ARMA-GARCH models, and by Boussama et al. (2011) for BEKK models. In the case of DCC models, a keystone is missing: a theory for inference has been proposed by Engle and Sheppard (2001), but their two stage estimation procedure is contingent on the underlying DCC process being strictly stationary and ergodic (see their Assumption A.2). The goal of this paper is to fill this gap.

\mds

After introducing some notations, we define DCC models at the beginning of Section~\ref{DCCrewriten}. They will be rewritten as
``almost linear'' Markov chains in Subsection~\ref{DCCMarkov}. The existence of strong and weak stationary solutions is stated in
Subsection~\ref{MainRes}. Subsection~\ref{Unicity} exhibits sufficient conditions to get their uniqueness. We discuss the tightness of our technical conditions from a qualitative standpoint in Section~\ref{My_discussion}. Proof of the propositions and theorems are detailed in the appendices.

\mds

\subsection{Notations}
\label{notations}
Consider an $(n,m)$ matrix $M=[m_{ij}]_{1\leq i
\leq n, 1\leq j \leq m}$.
\begin{itemize}
\item $M\geq 0$ (resp. $M>0$) means that all elements of $M$ are non-negative (resp. strictly positive), and $|M|=[|m_{ij}|]_{1\leq i
\leq n, 1\leq j \leq m}$.
\item If $n=m$, let the diagonal matrix $diag(M)=[m_{ij}\1 (i=j)]_{1\leq i
\leq m, 1\leq j \leq m}$ and the vector $Vecd(M)=[m_{ii}]_{1\leq i \leq m}$ in $\RR^m$.
\item If $n=m$ and $M$ is symmetric, $Vech(M)$ denotes the $m(m+1)/2=:m^*$ column vector whose components are read from $M$ column-wise and without redundancy.
To be formal, $Vech(M)=[\tilde m_{k}]_{1\leq k \leq
m^*}$, where $\tilde m_{k}=m_{ij}$ for the unique couple of
indices $(i,j)$ in $\{1,\ldots,m\}^2$, $i\geq j$ such that $ [m
+ (m-1) + \ldots + (m-j+2)]^+ + (i-j+1) =k$. This defines
a one-to-one mapping $\phi$ between the indices
$k\in\{1,\ldots,m^*\}$ and the pairs $(i,j)$, $i\geq j$, $1\leq
i,j \leq m$, i.e. $(i,j)=(\phi_1(k),\phi_2(k))=:\phi(k)$.
\item $\otimes$ denotes the usual Kronecker product, and
$M^{\otimes p} = M\otimes \ldots \otimes M$ ($p$ times). $\odot$ denotes the element-by-element product. If $v$ is a
vector in $\RR^n$, then $v\odot M = [v_i m_{ij}]_{1\leq i \leq
n,1\leq j\leq m}$.
\item We will consider several matrix norms, particularly
$$\|M\|_{\max} = \max_{1\leq i\leq n,1\leq j\leq m}|m_{ij}|,$$
and the spectral norm, defined for any squared matrix by
$$ \| M \|_s = \sup\{\sqrt{\lambda}\;| \; \lambda \;\text{is an eigenvalue of}\; M'M\}=\sup_\xx \frac{\|M\xx\|_2}{\| \xx \|_2}\cdot$$
Besides, we will consider any norm $\Nc$ for vectors, that is not the Euclidian norm $\|\cdot\|_2$. Then, we can define the norm $\|\cdot\|_{\Nc}$ for matrices by setting
$ \| M \|_{\Nc} = \sup_\xx  \Nc(M\xx)/\Nc(\xx).$ Note that $\|M\|_{\infty}= \max_{i} \sum_{j} |m_{ij}|$ when $\Nc(\xx)=\|\xx\|_{\infty}=\max_i |x_i|$.
\item $\rho(M)$ denotes the spectral radius of the squared matrix
$M$, i.e. the largest of the modulus of $M$'s eigenvalues. If
$M$ is positive semi-definite, then $\rho(M)= \|M\|_s$ and its smallest eigenvalue is
denoted by $\lambda_1(M)$.
\item For any column vector $z_t\in \RR^m$, we denote
$z_t=(z_{1,t},\ldots,z_{m,t})'$ and
$\vec{z}_t:=(z_{1,t}^2,\ldots,z_{m,t}^2)'$.
\item $e$ denotes a vector of ones, the dimension of which will be
implicit. $0_m$ (resp. $I_m$) denotes the $m\times m$ matrix of zeros (resp. identity
matrix). When the dimension of an identity matrix is not
specified, it will be denoted by $Id$.
\item If $M$ depends on $\xx\in A$, then $\sup_{\xx \in A} M(\xx)$ is the matrix
$[\sup_{\xx\in A} m_{ij}(\xx)]$.
\end{itemize}

\mds

\section{Dynamic Conditional Correlation models}
\label{DCCrewriten}
\subsection{The classical DCC specification}

Let us reiterate here the standard DCC model, as introduced in Engle (2002). Consider a stochastic process $(y_t)_{t\in\ZZ}$ in $\RR^m$, typically a vector of $m$ asset returns. The sigma field generated by the past information of this process until up to and including time $t-1$ is denoted by $\Ic_{t-1}$.

Modeling the expected returns of financial series is a problem per se,
that has generated a huge amount of literature. In this paper, our
focus will be on the dynamics of the conditional variance-covariance
of $y_t$ instead. Therefore, following current practice, we will assume we
can remove the conditional means of our returns. Let
$\mu_t(\theta)=E[y_t | \Ic_{t-1}]=: E_{t-1}[y_t]$ be the
conditional mean vector of $y_t$. It depends on a vector of
parameters $\theta\in\Theta$. We define a ``detrended'' series
$(z_t)_{t\in\ZZ}$ by $$ y_t = \mu_t(\theta) +z_t,\;\;
E_{t-1}[z_t]=0.$$ For convenience, the conditional mean
$\mu_t(\theta)$ is assumed to be measurable with respect to
$\sigma(z_{t-1},z_{t-2},\ldots)$. Therefore,
$\Ic_t=\sigma(y_t,y_{t-1},\ldots) = \sigma(z_t,z_{t-1},\ldots)$.

Let us denote by $H_t$ the variance-covariance
matrix of the $t$-observations, conditionally on $\Ic_{t-1}$:
$Var(y_t|\Ic_{t-1})=Var(z_t|\Ic_{t-1}):=H_t.$
As usual with DCC-type models, we split the variance-covariance
matrix $H_t$ between volatility terms on one side (in $D_t$), and
correlation coefficients on the other side (in $R_t$):
\begin{equation}
H_t=D_t^{1/2} R_t D_t^{1/2}, \;\;D_t=diag(h_{1,t},...,h_{m,t}),
\label{Htdef}
\end{equation}
where $h_{k,t}$ denotes the ``instantaneous variance'' of
the return $y_{k,t}$ (or $z_{k,t}$, equivalently), conditionally on
$\Ic_{t-1}$.
We assume GARCH-type models on every
margin, but with potential cross-effects between all these volatilities:
\begin{equation}
Vecd(D_t) = V_0 + \sum_{i=1}^r
A_i .Vecd(D_{t-i}) + \sum_{j=1}^s B_j. \vec{z}_{t-j},
\label{Dtdef}
\end{equation}
for some deterministic non-negative matrices $(A_i)_{i=1,\ldots,r}$ and
$(B_j)_{j=1,\ldots,s}$, and for a positive vector $V_0$ in $\RR^m$. We will set
$A_i:=[a^{(i)}_{k,l}]_{1\leq k,l \leq m}$, $i=1,\ldots,r$, and
$B_j:=[b^{(j)}_{k,l}]_{1\leq k,l \leq m}$, $j=1,\ldots,s$.

\mds

Let us introduce the vector of so-called ``standardized residuals''
$\varepsilon_t := D_t^{-1/2}z_t$. Obviously, $E_{t-1}[\varepsilon_t]=0$ and $E_{t-1}[\varepsilon_t \varepsilon_t']=R_t$.
We impose that $M^{1/2}$ is positive definite for any positive definite matrix $M$. In this case, the
square root of $R_t$ is uniquely defined: see Serre (2010), Theorem 6.1. This will be our convention throughout the article.

\mds

The dynamics of correlations are given by the traditional Dynamic Conditional Correlation specification:
\begin{equation}
R_t=diag(Q_t)^{-\frac{1}{2}} Q_t diag(Q_t)^{-\frac{1}{2}}, \label{RtDCC}
\end{equation}
where the sequence of matrices $(Q_t)_{t\in\ZZ}$ satisfies
\begin{equation}
 Q_t= W_0 + \sum_{k=1}^\nu M_k Q_{t-k} M_k' +
\sum_{l=1}^\mu N_l \eps_{t-l} \eps_{t-l}' N_l',
\label{Qtdef}
\end{equation}
for some deterministic matrices $(M_k)_{k=1,\ldots,\nu}$ and
$(N_l)_{l=1,\ldots,\mu}$, and for a positive definite constant
matrix $W_0$. Obviously, when such a sequence $(Q_t)_{t\geq -\nu}$ is initialized with $\nu$ non negative definite (possibly null) matrices, every $Q_t$, $ t\geq 0$, will be definite positive. In Theorem~\ref{thMoments}, we prove that a ``doubly infinite'' stationary sequence $(Q_t)_{t\in \ZZ}$ of definite positive matrices exists and which satisfies~(\ref{Qtdef}).

\mds

We will set $M_k:=[m^{(k)}_{p,q}]_{1\leq p,q \leq m}$, $k=1,\ldots,\nu$, and $N_l:=[n^{(l)}_{p,q}]_{1\leq p,q \leq m}$, $l=1,\ldots,\mu$. In practice, the positive matrix $W_0$ (or the constant vector $Vech(W_0)$ in $\RR^{m^*}$ equivalently) is a parameter that has to be estimated, most often during the first stage.

\mds

Aielli (2013) noticed that the estimation of the unknown matrix $W_0$ is not straightforward, because it cannot be deduced trivially from the unconditional correlation between the standardized residuals $\eps_t$. Therefore, he introduced a new variety of DCC-GARCH models (called cDCC), where~(\ref{Qtdef}) is replaced by
\begin{equation}
 Q_t= W_0 + \sum_{k=1}^\nu M_k Q_{t-k} M_k' +
\sum_{l=1}^\mu N_l diag(Q_{t-l})^{-1/2} \eps_{t-l} \eps_{t-l}' diag(Q_{t-l})^{-1/2} N_l'.
\label{QtdefAielli}
\end{equation}
Under this new assumption, cDCC can be seen as a particular BEKK model (Engle and Kroner, 1995). Therefore, Aielli obtained the
existence of strictly and/or weakly stationary solutions, applying the conditions of Boussama, Fuchs, and Stelzer (2011) on BEKK processes. Actually, Aielli's model~(\ref{QtdefAielli}) is a smart but not intuitive ``ad-hoc'' specification. Its main justification appears as essentially technical, to avoid the non-linear feature of Engle's original DCC model~(\ref{Qtdef}). Under the standard latter specification, DCC models can no longer be rewritten as BEKK models and other techniques have to be found. In this paper, we obtain similar results to Aielli (2013), but by keeping the original specification of DCC models and without relying on another surrounding family of processes.

\subsection{DCC as Markov chains}
\label{DCCMarkov}
Actually, it is possible to rewrite the previous DCC model as a Markov chain, that looks like an AR(1) process. This rewrite will become a crucial tool when studying stationary solutions hereafter. Set
\begin{equation}
X_t := (X_t^{(1)},X_t^{(2)},X_t^{(3)},X_t^{(4)})',
\label{defXt}
\end{equation}
where
$$ X_t^{(1)}:= (Vecd(D_t),\ldots, Vecd(D_{t-r+1}))',$$
$$ X_t^{(2)} := (\vec{z}_t,\ldots,\vec{z}_{t-s+1})',$$
$$ X_t^{(3)}:= (Vech(Q_t),\ldots, Vech(Q_{t-\nu+1}))',\;\text{and}$$
$$ X_t^{(4)} := (Vech(\eps_t\eps_t'),\ldots,Vech(\eps_{t-\mu+1}\eps_{t-\mu+1}'))'.$$
The dimensions of the four previous random vectors are $rm$, $sm$,
$\nu m^*$ and $\mu m^*$ respectively. Their sum, the dimension of
$X_t$, is denoted by $d$. With simple block matrix calculations, random matrices $(T_t)$ and a vector process
$(\zeta_t)$ exist, such that the dynamics of $X_t$, any solution of the DCC model, may be rewritten as
\begin{equation}
 X_t = T_t. X_{t-1}+\zeta_t,
 \label{defMarkov}
\end{equation}
for any $t$. We will write the block matrix $T_t:=[T_{ij,t}]_{1\leq i,j\leq 4}$ with convenient
random matrices $T_{ij,t}$.

\mds

Knowing~(\ref{defMarkov}), the underlying process $(X_t)$ can be
seen as a vectorial autoregressive of order one, but with random
matrix-coefficients $(T_t)$.
Let us detail the AR(1) form of~(\ref{defMarkov}):
\begin{itemize}
\item set $T_{1k,t}=0$ when $k=3,4$,
$$ T_{11,t}:=
\left[
\begin{matrix}
A_1 & A_2 & \cdots & \cdots & A_r \\
I_m & 0_m & \cdots &  \cdots & 0_m \\
0_m & I_m & 0_m &  & \vdots \\
\vdots & \ddots & \ddots & \ddots & \vdots \\
0_m & \cdots & 0_m & I_m & 0_m
\end{matrix}
\right],\; \text{and}\;\,
T_{12,t}:= \left[
\begin{matrix}
B_1 & B_2 & \cdots & \cdots & B_s \\
0_m & \cdots & \cdots &  \cdots & 0_m \\
\vdots &  &  &  & \vdots \\
\vdots &  &  &  & \vdots \\
0_m & \cdots & \cdots & \cdots & 0_m
\end{matrix}
\right]. $$

\item
We deduce from Equation~(\ref{Dtdef}) that
\begin{equation}
D_t \vec{\eps}_t  = \vec{\eps}_t \odot Vecd(D_t) =\vec{z}_t = \vec{\eps}_t \odot V_0 + \sum_{i=1}^r
\vec{\eps}_t \odot A_i .Vecd(D_{t-i}) + \sum_{j=1}^s \vec{\eps}_t \odot B_j .\vec{z}_{t-j}.
\label{Dtdef2}
\end{equation}
Let us set $T_{23,t}=T_{24,t}=0$,
$$ T_{21,t}:=
\left[
\begin{matrix}
\vec{\eps}_t \odot  A_1 & \vec{\eps}_t \odot  A_2 & \cdots & \cdots & \vec{\eps}_t \odot  A_r \\
0_m & \cdots & \cdots &  \cdots & 0_m \\
\vdots &  &  &  & \vdots \\
0_m & \cdots & \cdots & \cdots & 0_m
\end{matrix}
\right],\; \text{and} $$

$$
T_{22,t}:= \left[
\begin{matrix}
\vec{\eps}_t \odot  B_1 & \vec{\eps}_t \odot  B_2 & \cdots & \cdots & \vec{\eps}_t \odot  B_s \\
I_m & 0_m & \cdots &  \cdots & 0_m \\
0_m & I_m & 0_m &  & \vdots \\
\vdots & \ddots & \ddots & \ddots & \vdots \\
0_m & \cdots & 0_m & I_m & 0_m
\end{matrix}
\right] $$
\item
Clearly,  matrices $\tilde M_k$, $k=1,\ldots,\nu$, exist such that
$$ Vech(M_k Q_{t-k} M_k')=\tilde{M}_k .Vech(Q_{t-k}).$$
Similarly, matrices $\tilde N_l$, $l=1,\ldots,\mu$, exist such that
$$ Vech(N_l \eps_{t-l} \eps_{t-l}' N_l')=\tilde{N}_l .Vech(\eps_{t-l}.\eps_{t-l}').$$

It is possible to explicitly write the previous matrices $\tilde M_k$ and $\tilde N_l$.s
Indeed, with the notations of Subsection~\ref{notations}, $ \tilde M_k = [\tilde
m_{u,v}^{(k)}]_{1\leq u,v \leq m^*}$ where
$$ \tilde m_{u,v}^{(k)} = m^{(k)}_{\phi_1(u),\phi_1(v)}m^{(k)}_{\phi_2(u),\phi_2(v)}.$$
Similarly,
$ \tilde N_l = [\tilde
n_{u,v}^{(l)}]_{1\leq u,v \leq m^*}$ and
$ \tilde n_{u,v}^{(l)} = n^{(l)}_{\phi_1(u),\phi_1(v)}m^{(l)}_{\phi_2(u),\phi_2(v)} .$
Then, set $T_{31,t}=T_{32,t}=0$,
$$ T_{33,t}:=
\left[
\begin{matrix}
\tilde{M}_1 & \tilde{M}_2 & \cdots & \cdots & \tilde{M}_\nu \\
I_{m^*} & 0_{m^*} & \cdots &  \cdots & 0_{m^*} \\
0_{m^*} & I_{m^*} & 0_{m^*} &  & \vdots \\
\vdots & \ddots & \ddots & \ddots & \vdots \\
0_{m^*} & \cdots & 0_{m^*} & I_{m^*} & 0_{m^*}
\end{matrix}
\right] ,\; \text{and}\;
T_{34,t}:=
\left[
\begin{matrix}
\tilde{N}_1 & \tilde{N}_2 & \cdots & \cdots & \tilde{N}_\mu \\
0_{m^*} & \cdots & \cdots & \cdots & 0_{m^*}  \\
\vdots &  &  &  & \vdots \\
0_{m^*} & \cdots & \cdots & \cdots & 0_{m^*}
\end{matrix}
\right]. $$
\item
$T_{4k,t}=0$, $k=1,2,3$, and define the $\mu m^* \times \mu m^*$ matrix
$$T_{44,t}:=
\left[
\begin{matrix}
0_{m^*} & 0_{m^*} & \cdots & \cdots & 0_{m^*} \\
I_{m^*} & 0_{m^*} & \cdots &  \cdots & 0_{m^*} \\
0_{m^*} & I_{m^*} & 0_{m^*} &  & \vdots \\
\vdots & \ddots & \ddots & \ddots & \vdots \\
0_{m^*} & \cdots & 0_{m^*} & I_{m^*} & 0_{m^*}
\end{matrix}
\right] .$$
\end{itemize}
Moreover, rewrite
$\zeta_t=(\zeta_t^{(1)},\zeta_t^{(2)},\zeta_t^{(3)},\zeta_t^{(4)}),$
where, with obvious sizes, these vectors are
$$ \zeta_t^{(1)}=(V_0,0_m,\ldots,0_m)',\;\; \zeta_t^{(2)}=(\vec{\eps}_t \odot V_0 ,0_m,\ldots,0_m)',$$
$$ \zeta_t^{(3)}=(Vech(W_0),0_{m^*},\ldots,0_{m^*})',\;\text{and}\;\; \zeta_t^{(4)}=(Vech(\eps_t\eps_t'),0_{m^*},\ldots,0_{m^*})'.$$

\mds

Since $D_t$, $Q_t$ and $R_t$ are
$\Ic_{t-1}-$measurable, it is easy to see that the filtration induced by the observations is the natural filtration of $(X_t)$:
$\sigma(X_t,X_{t-1},\ldots)=\Ic_t$ for all $t$. From now on, we consider $(\Ic_t)$ as the filtration that is generated by $(X_t)$. This means that $E_{t-1}[Z]=E[Z | \Ic_{t-1}]=E[Z | X_{t-1},X_{t-2},\ldots]$, for any random vector $Z$. And a process $(Z_t)$ is said to be (one-order, implicitly) $\Ic$-Markov if the law of $Z_t$ given $\Ic_{t-1}$ is the
law of $Z_t$ given $X_{t-1}$.

\mds

Intuitively, the sequence $(X_t)$ is $\Ic$-Markov because it is the
case for the processes $(\zeta_t)$ and $(T_t)$ themselves. To prove this formally, we need an assumption concerning the data generating process (DGP) of $(z_t)$.

\mds

Let us define the $t$-vector of innovations by
\begin{equation}
\eta_t := R_t^{-1/2} \varepsilon_t = R_t^{-1/2} D_t^{-1/2} z_t.
\label{defEta}
\end{equation}
Note that $E_{t-1}[\eta_t]=0$ and $E_{t-1}[\eta_t\eta_t']=I_m$ by construction.
The definition of
these innovations implies that, for every $t$,
$\sigma(\eta_j, j\leq t) \subset \sigma(\eps_j, j\leq t) \subset \Ic_t .$
Nonetheless, we will not establish whether there are equalities between
the latter filtrations. Technically speaking, this would be
equivalent to stating the invertibility of the underlying process.

\mds

{\it\bf Assumption A0:}
$(\eta_t)_{t\in \ZZ}$ possesses the Markov property with respect to the filtration $\Ic$. In particular, $E[\eta_t | \Ic_{t-1}]=
E[\eta_t |X_{t-1}]$ for every $t$.

\mds

Obviously, the latter assumption is satisfied if $(\eta_t)_{t\in \ZZ}$ is a sequence of identically distributed and mutually
independent random vectors, with $E[\eta_t]=0$ and $E[\eta_t \eta_t']=I_m$.
For instance, if the random vectors $\eta_t$ are standardized Gaussian and mutually independent, the process $(z_t)_{t\in\ZZ}$ is conditionally Gaussian, a standard case in practice.

\mds

\begin{proposition}
\label{prop_Markov}
Under A0, the process $(X_t)$ is Markov of order one with respect to its natural filtration.
\end{proposition}
See the proof in the appendix.

\begin{remark}
It would be possible to define a slightly different DGP for the DCC model above: consider a $\Ic$-martingale difference (i.i.d., for instance) sequence $(e_t)_{t\in\ZZ}$ and set
$z_t = H_t^{1/2} e_t$, $E_{t-1}[e_t]=0$, $E_{t-1}[e_t e'_t]=I_m$. In the latter case, the process $(\eta_t)$ above would be defined by $\eta_t = R_t^{-1/2} D_t^{-1/2} H_t^{1/2} e_t$. Then, it is easy to check that $(\eta_t)$ is $\Ic$-Markov and
is a martingale difference. In other words, A0 would apply in such circumstances.
\end{remark}

\section{Stationarity of DCC models}

\subsection{Existence of stationary DCC solutions}
\label{MainRes}

The AR dynamics of $X_t$ were defined above thanks to $T_t$ and $\zeta_t$, which will be stochastic only through $\eps_t$, i.e. through the $t$-innovation $\eta_t$ and the $\Ic_{t-1}$-measurable matrix $R_t$. This creates a major difficulty in proving the existence of stationary solutions. In particular, this means that $T_t$ depends on some components of $X_t$. Therefore, it will be difficult to find explicit expressions
like $X_t = f(\eta_t,\eta_{t-1},\ldots)$ for some deterministic and measurable function $f$, because the link between $T_t$ and the past
innovations (or observations) is highly non-linear.

\mds

To obtain the existence of stationary solutions in the previous DCC model, we will invoke Tweedie's (1988) criterion. The latter result will provide the existence of an invariant probability measure for the Markov chain defined by~(\ref{defMarkov}). sThis technique has already been used in several papers in econometrics, notably Ling and McAleer (2003) or Ling (1999).

\mds

To get the stationarity conditions of $(z_t)$, we have to control the magnitude of the  random matrix $T_t$, which depends on the random variables $\eps_{kt}^2$, $k=1,\ldots,m$. The mean of the latter variables is one, but they are not independent. This is in contrast with Ling and McAleer (2003). Moreover, unfortunately, the joint law of $\vec{\eps}_t$ is a function of $R_t$, i.e. a function of $X_{t-1}$. That is why we need
the following condition.s

\mds

{\it\bf Assumption E1:} For some $p\geq 1$, $E[\| \eta_t
\|^{2p}]<\infty$ and $ \rho\left(T^*\right)<1$, where
$$ T^*:=\sup_{\xx\in \RR^d} E[|T_t^{\otimes p}| \; | \; X_{t-1}=\xx ].$$

\mds

We reiterate that $T_t$ depends on $\vec{\eps}_t$, that $\eps_t=R_t^{1/2} \eta_t$, and that the components of $\eta_t$ are uncorrelated. As such, the coefficients of $ T^*$ are finite because all of the coefficients of $R_t$ are less than one (in absolute values). When there are no correlation dynamics, the matrices $M_k$ and $N_l$ are zero and we recover CCC models. In the latter case, our Assumption E1 is reduced to the main assumption of Ling and McAleer (Theorem 2.2) that was stated for vectorial ARMA-GARCH models.

\mds

{\it\bf Assumption E2:}
The law of $\eta_t$ given that $X_{t-1}=\xx$ is absolutely continuous with respect to the Lebesgue measure, and its density is denoted by $f_{\eta_t}(\cdot |\xx)$, for every $\xx \in \RR^d$ and $t$.
The function $\xx \mapsto f_{\eta_t}(\eta | \xx)$ is continuous for every $\eta\in \RR^m$ and $t$.
There exists an integrable function $H$ s.t. $ \sup_t\sup_{\xx\in \RR^d} f_{\eta_t}(\eta |\xx) \leq H(\eta)$
for every $\eta\in \RR^m$.
Moreover, $\sup_t E[\| \eta_t \|^{2p} \, | X_{t-1}=\xx] \leq \bar h(\|\xx \|)$, for some function $\bar h$ that satisfies
$\lim_{v\rightarrow +\infty} \bar h(v)/v^\gamma =0$ for every $\gamma >0$.

\mds

The latter technical assumption is trivially satisfied when $(\eta_t)$ is an i.i.d. sequence of random vectors s.t. $E[\| \eta_t \|^{2p}]<+\infty$.
Otherwise, E2 provides some constraints insofar as the law of $\eta_t$ depends on the past values of the DCC process.
Similar conditions may appear in the literature about the non-parametric estimation of conditional expectations.
However, most of them relate to the boundedness of $\bar h$ and/or its derivatives (as in Assumption 3 in Newey 1997, for instance), or to the moments of $\bar h$ (as Assumption 1 in Donald et al. 2003, for instance). Clearly, E2 is weaker than such assumptions.

\mds

\begin{theorem}
\label{thMoments} Under the assumptions A0 and E1-E2, the process $(z_t,D_t,R_t)$
as defined by Equations~(\ref{Htdef}),~(\ref{Dtdef}),~(\ref{RtDCC})
and~(\ref{Qtdef}), possesses a strictly stationary solution. The
latter process is measurable with respect to the $\sigma$-field $\Ic$ induced
by the observations. Moreover, the $2p$-th moments of a solution $(z_t)$ are finite.
\end{theorem}

{\bf Example 1:} In practice and for the sake of parsimony, it is usual to assume diagonal-type DCC models, where all the
parameter matrices are diagonal, assuming no ``cross-effects''
in terms of volatilities and/or correlations. This means the
non-negative real numbers $a^{(i)}_u$, $b^{(j)}_u$, $m^{(k)}_u$ and $n^{(l)}_u$,
$u=1,\ldots,m$, exist such that
$$ A_i = diag(a^{(i)}_1,\ldots,a^{(i)}_m),\, i=1,\ldots,r,\; B_j=diag(b^{(j)}_1,\ldots,b^{(j)}_m),\, j=1,\ldots,s,$$
$$ M_k = diag(m^{(k)}_1,\ldots,m^{(k)}_{m^*}),\, k=1,\ldots,\nu, \; N_l=diag(n^{(l)}_1,\ldots,n^{(l)}_{m^*}),\, l=1,\ldots,\mu.$$
The associated matrices $\tilde M_k$ and $\tilde N_l$ are also diagonal.
Set $\tilde M_k = diag(\tilde m^{(k)}_l)_{1\leq l \leq m^*}$, and check that $\tilde m^{(k)}_l=m_{\phi_1(l)}^{(k)} m_{\phi_2(l)}^{(k)}$.
Now, let us specify the previous Assumption E1 when $p=1$.

\mds

Since $E[\vec{\eps}_{kt}\; | \; X_{t-1}=\xx ]=1$ for every index $k$,
$T^*$ is simply $|T_t|$, replacing $\vec{\eps}_t$ by one. Denote
by $P^*$ the characteristic polynomial of $T^*$, i.e.
$P^*(\lambda)=\text{Det}(T^* - \lambda Id)$. It can be seen easily
that two polynomials $P_1^*$ and $P_2^*$ s.t.
$ P^*(\lambda)= P_1^*(\lambda)  P_2^*(\lambda)$ exist. Here, $P_1^*$ denotes the
characteristic polynomial of the block-matrix $[|T_{ij,t}|]_{1,\leq
i,j\leq 2}$, replacing $\vec{\eps}_t$ by one. $P_2^*$ is the
characteristic polynomial of the previous matrix $|T_{33,t}|$. Tedious, but relatively uncomplicated,
algebraic calculations provide
$$ P_1^*(\lambda) = \pm \lambda^{\pi_1} \Prod_{k=1}^m \left(\sum_{i=1}^r a_k^{(i)}\lambda^{r+s-i} + \sum_{j=1}^s
b_k^{(j)} \lambda^{r+s-j} -\lambda^{r+s}\right),$$
$$ P_2^*(\lambda) = \pm \lambda^{\pi_2} \Prod_{l=1}^{m^*} \left(\sum_{k=1}^\nu \tilde m^{(k)}_l \lambda^{\nu-k} -\lambda^{\nu}\right),$$
for some integers $\pi_1$ and $\pi_2$.
Let $\lambda_0$ be a non-zero root of $P^*$. If $\lambda_0$ is a root of $P_1^*$ then there exists an index $k\in\{1,\ldots,m\}$ such that
$ \sum_{i=1}^r a_k^{(i)}\lambda_0^{r+s-i} + \sum_{j=1}^s
b_k^{(j)} \lambda_0^{r+s-j} =\lambda_0^{r+s} .$
If $|\lambda_0|\geq 1$, this implies
$ 1 \leq \sum_{i=1}^r a^{(i)}_k  + \sum_{j=1}^s b^{(j)}_k.$

On the other side and similarly, if $\lambda_0$ is a root of $P_2^*$ and if $|\lambda_0|\geq 1$, then there exists $l\in\{1,\ldots,m^*\}$ s.t.
$ 1 \leq \sum_{k=1}^\nu |\tilde m^{(k)}_l| .$
In other words, a sufficient condition to fulfill Assumption E1 is
\begin{equation}
 \sup_{k=1,\ldots,m} \sum_{i=1}^r a^{(i)}_k  + \sum_{j=1}^s b^{(j)}_k <1,\;\; \text{and}\;\;
 \sup_{l=1,\ldots,m^*}\sum_{k=1}^\nu |\tilde m^{(k)}_l| <1 .
 \label{Th1DiagDCCCond}
 \end{equation}

\mds

Nonetheless, to apply Theorem~\ref{thMoments} in the general case, it may be hard to check the condition
on the spectral radius of $T^*$. This is due to
the analytical complexity of $T_t^{\otimes p}$, $p>1$, or to the calculation of its eigenvalues, even when $p=1$. In the next theorem,
we provide more explicit conditions in the case $p=1$, i.e. so that the second-order moments of $(z_t)$ are finite. These conditions ensure that E1 will be satisfied.
In other words, the conditions will be stronger than E1, but they may be more practical.
Indeed, it is often important to obtain sufficient conditions that can be written explicitly in terms of the model parameters, for instance for inference purposes (e.g. the optimization stage to get QML estimates).

\mds

Let us consider $\Nc$ (resp. $\Nc^*$) an arbitrary norm for vectors in $\RR^m$ (resp. $\RR^{m^*}$). Denote by $\|\cdot \|_{\Nc}$
and $\|\cdot \|_{\Nc^*}$ the associated norms for matrices.
\begin{theorem} If
\label{thSimple}
\begin{equation}
 \sum_{i=1}^r \| A_i \|_{\Nc} + \sum_{j=1}^s  \| B_j \|_{\Nc} <1,\;\; \text{and}
\label{cond1ThSimple}
\end{equation}
\begin{equation}
\sum_{k=1}^\nu  \| \tilde M_k \|_{\Nc^*} <1,
\label{cond2ThSimple}
\end{equation}
then Assumption E1 is satisfied with $p=1$.
\end{theorem}
Note that the conditions of Theorem~\ref{thSimple} do not depend on
the matrices $N_l$, $l=1,\ldots,\mu$, which is a relatively unexpected result.
Once they are satisfied, and under A0 and E2, Theorem~\ref{thMoments} applies.

\mds

By choosing $\Nc$ as the maximum norm for vectors, it can easily be checked that $\|A_i\|_{\Nc}=\sup_{p=1,\ldots,m} \sum_{q=1}^m a^{(i)}_{p,q}$, and similarly with the matrices $B_j$.
Alternatively, we can choose $\Nc(\xx)=\|x\|_2$, that induces the spectral norm $\|A_i \|_{\Nc}=\|A_i \|_s$. Obviously,
we can choose these norms for $\Nc^*$ and the matrices $\tilde M_k$.

%

\mds

It is often of value to assume that the Markov chain is initialized at $t=0$ by drawing $X_0$ following its stationary law. Introducing the filtration $\Ic_t^*:=\sigma(X_0,z_1,\ldots,z_t)$, we can easily see that the DCC solution is now measurable with respect to the
$\sigma$-field induced by the innovations and the initial value, because, whenever $t>0$,
$$ \sigma(X_0,z_1,\ldots,z_t)=\sigma(X_0,\eps_1,\ldots,\eps_t)=\sigma(X_0,\eta_1,\ldots,\eta_t).$$

\mds

{\bf Example 1 (Continued):} Consider a diagonal-type DCC model and maximum norms for vectors.
In this case, the condition~(\ref{cond1ThSimple}) becomes
$$ \sum_{i=1}^r \sup_{l=1,\ldots,m} a_{l}^{(i)} + \sum_{j=1}^s \sup_{l=1,\ldots,m} b_{l}^{(j)} <1,$$
and the condition~(\ref{cond2ThSimple}) is
$\sum_{k=1}^\nu  \sup_{l=1,\ldots,m^*}|\tilde m^{(k)}_l| <1.$
These two conditions are stronger than~(\ref{Th1DiagDCCCond}), as expected.

\mds

{\bf Example 2:}
To reduce the number of free parameters even further, scalar-DCC models are often introduced. In
this case, all the unknown matrices are simply products of a
scalar and an identity matrix:
$$ A_i = a^{(i)}I_m,\, i=1,\ldots,r,\; B_j=b^{(j)}I_m,\, j=1,\ldots,s,$$
$$ M_k = m^{(k)}I_m,\, k=1,\ldots,\nu, \; N_l=n^{(l)}I_m,\, l=1,\ldots,\mu.$$
Such models are very popular, because they allow the number of free parameters to be drastically reduced. With obvious notations,
the conditions of Theorem~\ref{thMoments} and~\ref{thSimple} are the same as above:
\begin{equation}
 \sum_{i=1}^r a^{(i)} + \sum_{j=1}^s b^{(j)} <1,\; \text{and}\; \sum_{k=1}^\nu   |m^{(k)} |^2 <1.
\label{scalarDCCcond}
\end{equation}
In passing, we recover the usual (second-order and strict)
conditions of stationarity for GARCH-type models:
$$0\leq a^{(i)},b^{(j)}\leq 1,\;\text{and}\;\;\sum_{i=1}^r a^{(i)} + \sum_{j=1}^s b^{(j)} <1.$$

\mds

\subsection{Uniqueness of stationary DCC solutions}
\label{Unicity}

Even if stationary solutions of the DCC model do exist, we are not initially sure {\it a priori} that they are unique. Besides its theoretical interest, this problem has practical implications. For instance, for any process, the convergence of simulated trajectories towards the same stationary law, independently of the initialization stage, is a desirable feature. Moreover, the uniqueness of invariant measures of a Markov process implies the ergodicity of the stationary solution (see Douc et al. 2014, Corollary 7.17). This is particularly important for inference purposes. Indeed, the estimation of DCC models is typically based on M-estimates (Quasi Maximum Likelihood, for instance). These techniques rely heavily on uniform Laws of Large Numbers, that are most often deduced from the ergodicity of the process. The conditions for identifiability and consistency rely on some expectations with respect to the underlying invariant measure of the given stationary process. If, for a given set of parameters, several invariant measures exist, then it becomes difficult to check such conditions. Finally, with several underlying invariant measures, we cannot exclude the possibility of switches from one stationary trajectory to another, disturbing the econometric analysis (stationarity tests, statistical uncertainty around estimates, etc).

\mds

Unfortunately, this uniqueness is not given ``for free'' by Tweedie's Lemma~\ref{TweedieLemma}. Moreover, the usual arguments concerning
the uniqueness of stationary GARCH-type solutions do not apply here. Indeed, under the Markov-chain specification given by Equation~(\ref{defMarkov}), the matrix $T_t$ is itself a function of the random vector $X_t$ through the $\vec{\eps}_t$ factors. This is a
major difference with the CCC case, and we need to find another strategy. In this section, we provide some uniqueness results under some more or less restrictive assumptions.

\mds

Now, we will consider only stationary solutions of the DCC model, as given in Section~\ref{MainRes}.
We know that such solutions exist under the (sufficient) conditions of Theorem~\ref{thMoments} or~\ref{thSimple}, but it is not necessary to impose
such conditions from the outset. Obviously, we will need other technical assumptions.

\mds

{\it\bf Assumption U0:}
The sequence of innovations $(\eta_t)_{t\in \ZZ}$ is highly stationary and ergodic.

\mds

{\it\bf Assumption U1:} $\| T_{33} \|_s<1$.

\mds

The matrix $T_{33}$ has been introduced in
Subsection~\ref{DCCMarkov}, under the name $T_{33,t}$. Since
$T_{33,t}$ does not depend on time, we have removed the index $t$ here.

\mds

Thanks to the latter assumptions, we will be able to bound
$\|Q_t\|_{\max}$ from above by a stationary process $(q_t)$, and from below by a constant. Moreover, $\lambda_1(Q_t)$ will be bounded from
below. These tools will be crucial in proving the uniqueness of
stationary DCC solutions. To do so, let us introduce some intermediate quantities.
\begin{itemize}
\item The process $(q_t)$, defined by
$$q_t :=\frac{\| Vech(W_0) \|_2}{1- \| T_{33}\|_s} +  \sqrt{\frac{m^3(m+1)}{2}} \sum_{l=1}^\mu \| \tilde N_l\|_{s}\xi_{t-l},$$
where
$\xi_t := \sum_{k=0}^{+\infty} \| T_{33}\|_s^k  \|\eta_{t-k}\|_2^2.$
\item The constants $C_\lambda :=\lambda_1(W_0)$ and $C_q:=\min_{i=1,\ldots,m}(W_0)_{ii}$.
\item The constants
$$C^*_\lambda := \frac{ \lambda_1(W_0)}{1- \sum_{k=1}^\nu (m^{(k)})^2},\; \text{and} \;
C_q^*:=\frac{\min_{i=1,\ldots,m}(W_0)_{ii}}{1- \sum_{k=1}^\nu (m^{(k)})^2}\cdot$$
\item
$\kappa = \max(\nu,\mu)$ and, for every $j=1,\ldots,\kappa$, set
$$ \beta_{j,t}:= \1 ( j\leq \nu) \| M_j \|_s^2 + \1(j\leq \mu) \| N_j \|_s^2
\frac{4(2m+1)m^{1/2} }{\sqrt{C_\lambda}C_q} \|\eta_t \|_2^2
\sqrt{q_t}.$$
\end{itemize}

\mds

Let $N_t^*$ be the $(\kappa,\kappa)$-squared random matrix
$$ N_t^* :=
\left[
\begin{matrix}
\beta_{1,t} & \beta_{2,t} & \cdots & \cdots & \beta_{\kappa,t} \\
1 & 0 & \cdots &  \cdots & 0 \\
0 & 1 & 0 &  & \vdots \\
\vdots & \ddots & \ddots & \ddots & \vdots \\
0 & \cdots & 0 & 1 & 0
\end{matrix}
\right].$$

Note that the sequences $(\xi_t)$, $(q_t)$ and $(N_t^*)$ are stationary and ergodic because any $\xi_t$, $q_t$ or $N_t^*$ is a measurable function of the innovations $(\eta_t)$ that are stationary and ergodic under Assumption U0.

\mds

{\it\bf Assumption U2:} $E[\ln^+ \| N_t^*\|]<\infty$ and the top Lyapunov exponent of the sequence $(N_t^*)$, defined by
$\gamma_N:=\lim_{t\rightarrow +\infty} t^{-1}E[\ln (\| N_1^*N_2^*\ldots N_t^*\|) ]$, is strictly negative.

\mds

Such conditions are standard in the GARCH
literature (see Francq and Zako\"{\i}an, 2010, Section 2.2.2. for instance). Note that $\gamma_N \leq E[\ln \| N_1^*\|]$ for any norm $\|\cdot \|$.

\mds

Actually, the technical assumptions U1-U2 above will ensure the uniqueness of $(\eps_t)$, $(Q_t)$ and $(R_t)$ only. To get the uniqueness of $(D_t)$ and then of $(z_t)$ itself, we need a last assumption: with the notations of
Subsection~\ref{DCCMarkov}, set
$$ \bar{T}_t := \left[
\begin{array}{cc}
T_{11,t} & T_{12,t} \\
T_{21,t} & T_{22,t}
\end{array}
\right],\;
\text{and}\; \bar{T}^* = E[\bar T_t].$$
Note that $\bar{T}^*$ does not depend on any particular sequence $(\eps_t)$ nor $t$, because $E[\eps^2_{kt}]=1$ for every $k$.

\mds

{\it\bf Assumption U3:} $\rho(\bar T^*)<1$

\mds

\begin{theorem}
\label{ThUnicity} Under A0 and U0-U3, a strictly stationary
solution of the DCC model is unique and ergodic, given a sequence $(\eta_t)$.
\end{theorem}

\mds

The latter result can be strengthened in the following particular case, which is commonly encountered in the literature.

{\it\bf Assumption U4:} The underlying DCC model is ``partially''
scalar, i.e. scalars $m^{(k)}$ exist such that $M_k = m^{(k)}
I_m$ for all $k=1,\ldots,\nu$. Moreover, $\rho(M^*)<1$ by setting
$$ M^* :=
\left[
\begin{matrix}
(m^{(1)})^2 & (m^{(2)})^2 & \cdots & \cdots & (m^{(\nu)})^2 \\
1 & 0 & \cdots &  \cdots & 0 \\
0 & 1 & 0 &  & \vdots \\
\vdots & \ddots & \ddots & \ddots & \vdots \\
0 & \cdots & 0 & 1 & 0
\end{matrix}
\right].$$

\mds

Obviously, U4 is not mandatory to get our uniqueness result, even
if it allows the technical condition U2 to be weakened most often, by lowering the $\beta_{j,t}$ terms. In every
case, this ``partially'' scalar case encompasses the common
practice of scalar DCC (or scalar multivariate GARCH) models.

\begin{corollary}
\label{ThUnicityPL}
Under A0 and U0-U4, a strictly stationary
solution of the DCC model is unique and ergodic, given a sequence $(\eta_t)$, replacing $C_\lambda$ (resp. $C_q$) by
$C^*_\lambda$ (resp. $C_q^*$) in U2.
\end{corollary}

\mds

{\bf Example 2 (Continued):}
In the case of scalar DCC models of order one, it is easy to specify the conditions above. Here, $r=s=\nu=\mu=1$,
$$ A_1 = a^{(1)} I_m, \; B_1=b^{(1)} I_m, \; M_1 = m^{(1)}I_m, \; N_1=n^{(1)}I_m.$$
Assumptions U1 and U4 are equivalent and mean $|m^{(1)}|<1$.
Assumption U4 is fulfilled if $E[\ln \| N_1^*\|_{\max}]<0$, or if
\begin{equation}
\label{Ex2Unicity}
 E\left[ \ln\left( (m^{(1)})^2 +  (n^{(1)})^2
\frac{4(2m+1)m^{1/2} }{\sqrt{C_\lambda}C_q} \|\eta_t \|_2^2
\sqrt{q_t} \right) \right] <0.
\end{equation}
This expectation could be easily evaluated by simulation, by noting that $\eta_t$ and $q_t$ are independent.
Finally,
$$  \bar T^* = \left[
\begin{array}{cc}
a^{(1)} & b^{(1)} \\
a^{(1)} & b^{(1)}
\end{array}
\right] \otimes I_m
     .$$
Through elementary algebra, it can checked that the characteristic function of $\bar T^*$ is the function $x\mapsto (-x)^m(a^{(1)}+  b^{(1)} - x)^m$. Then Assumption U3 means $a^{(1)}+  b^{(1)} <1$.
Therefore, as expected, the conditions required for stationary DCC solutions to be unique are more demanding than for them to simply exist, due to U2. Generally, the latter condition will be fulfilled more easily if $\sup_l \|N_l \|_s$ is ``a lot smaller'' than one, if $m$ is not ``too large'', and if the tails of $\eta_t$ are not ``too heavy''.

\mds

\section{Discussion and practical considerations}
\label{My_discussion}
Now, let us discuss the sufficient conditions to obtain the existence of stationary DCC solutions, as given in Theorems~\ref{thMoments} and~\ref{thSimple}.
The most explicit ones are~(\ref{cond1ThSimple}) and~(\ref{cond2ThSimple}). Since scalar DCC models are by far the most commonly used models in the literature, we focus on the conditions of the previous examples 1 and 2 above, particularly~(\ref{scalarDCCcond}). In the latter case, the conditions on the coefficients of the volatility process are those generally applied in the univariate GARCH literature. It can be proved they are necessary and sufficient for the existence of second-order and strictly stationary GARCH solutions (see Francq and Zako\"{\i}an, 2010, Theorem 2.6 and Remark 2.6). More interestingly, we can check empirically how tight the (now) new conditions on the coefficients of the $(Q_t)$ process are. In other words, in Example 2, is the constraint
$ \sum_{k=1}^\nu   |m^{(k)} |^2 <1 $
close to a necessary and sufficient condition for generating stationary trajectories of $(z_t)$, $(R_t)$ and/or $(Q_t)$s?

\mds

For illustrative purposes, we have considered a very simple bivariate scalar DCC model of order one, given by
$$ h_{k,t}= v_{0} + a\, h_{k,t-1} + b \,\eps_{k,t}^2, \; k=1,2,$$
$$ v_0 =  1/4,\, a=0.8 ,\, b=0.1,$$
$$ Q_t = W_0 + (m^{(1)})^2 Q_{t-1} + (n^{(1)})^2 \eps_{t-1} \eps_{t-1}',$$
$$ W_0 = I_2/2 +ee'/2,\, e=[1,1]', $$
with our notations. The latter process is generated by i.i.d. innovations $(\eta_t)$ that are independent standard bivariate Gaussian vectors. We initialize
the process at $t=0$ with $Q_0=R_0=I_2$ and $h_{k,0}=1/2$, $k=1,2$.

\mds

In the paper, we have stated theoretically that the values of the coefficients of the matrices $N_l$, $l=1,\ldots, \mu$ (or the coefficients $n^{(l)}$ in the scalar case) do not matter to obtain the existence of stationary DCC solutions. We have verified this fairly counter-intuitive fact empirically: with the model above, different coefficients $n^{(1)}$
do not seem to modify the shape of the trajectories we generate, independently of the other parameters. Therefore, our experiments will lead with a fixed value $n^{(1)}=\sqrt{3}$.
 Note that this value is larger than one and this could be seen ``naively''  as a source of non-stationarity.

\mds

\begin{figure}[t!]
    \centering
    \begin{subfigure}[b]{1.1\textwidth}
        \centering
        \includegraphics[width=1.0\textwidth]{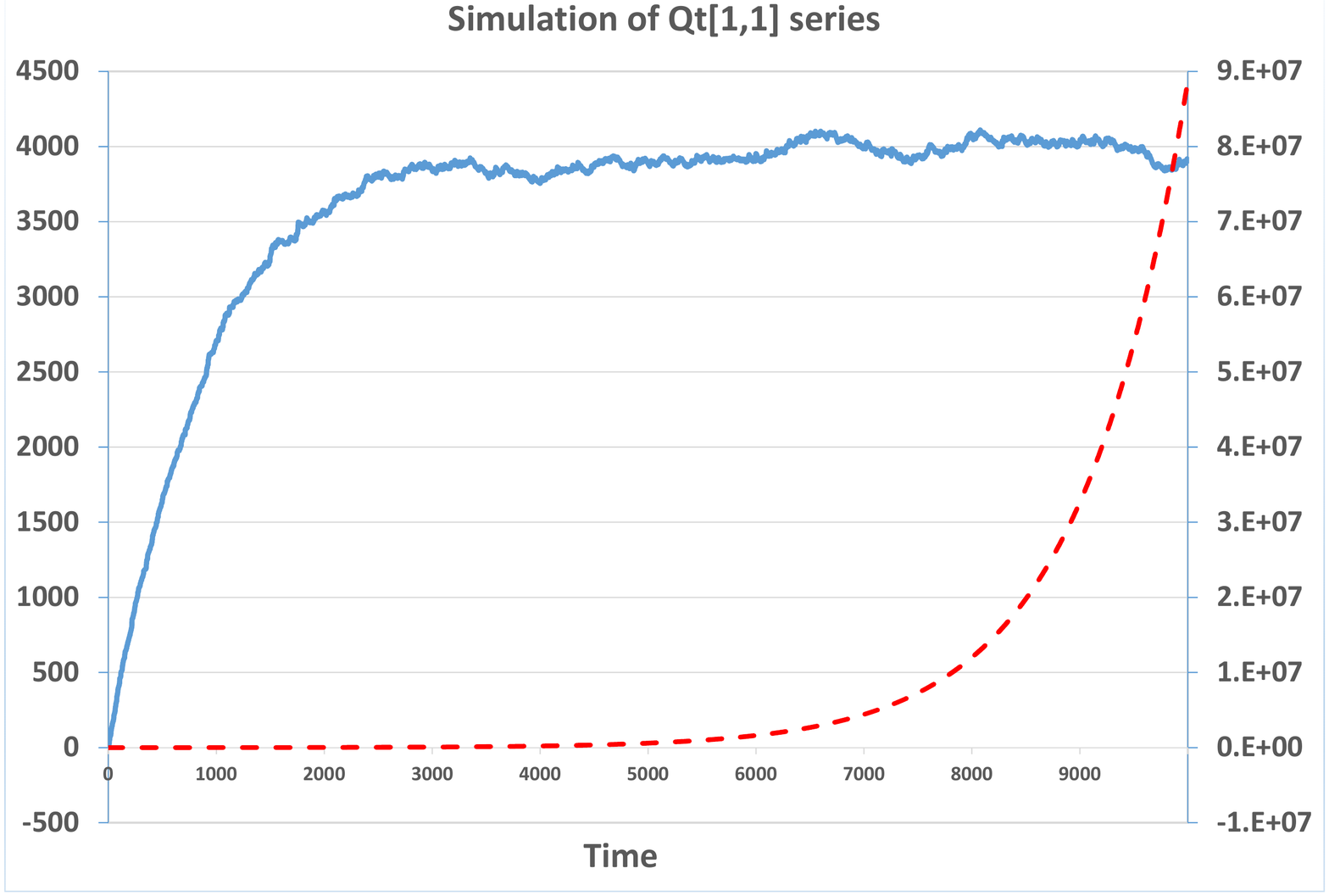}
    \end{subfigure}%
    \\
    \begin{subfigure}[b]{1.1\textwidth}
        \centering
        \includegraphics[width=1.0\textwidth]{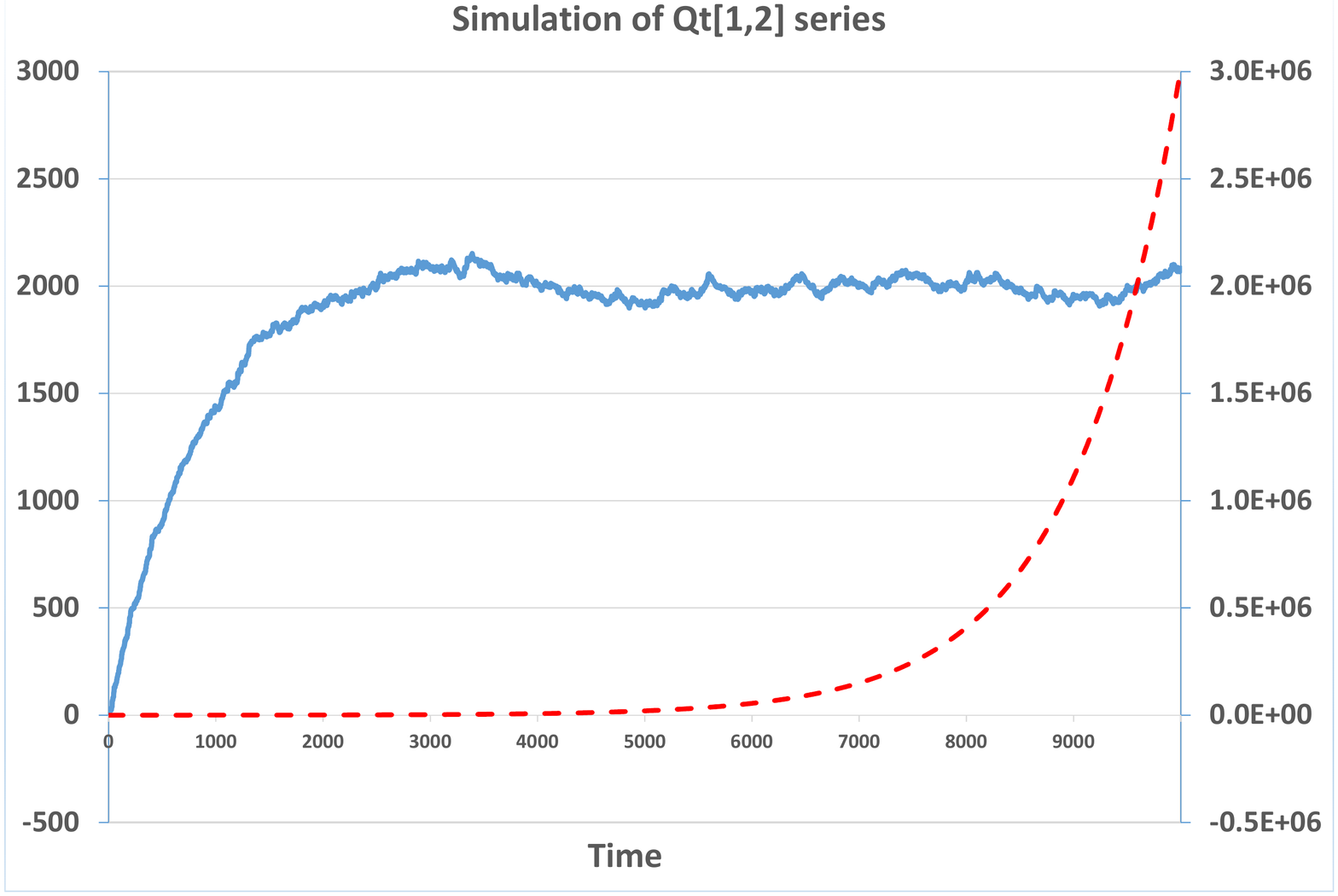}
    \end{subfigure}
\caption{Some simulated trajectories of $Q_t[1,1]$ (top) and $Q_t[1,2]$ (bottom) when $m^{(1)}=\sqrt{0.999}$ (solid line, left axis) or $m^{(1)}=\sqrt{1.001}$ (dashed line, right axis).} \label{QtFig}
\end{figure}

\begin{figure}[t!]
    \centering
        \includegraphics[width=1.0\textwidth]{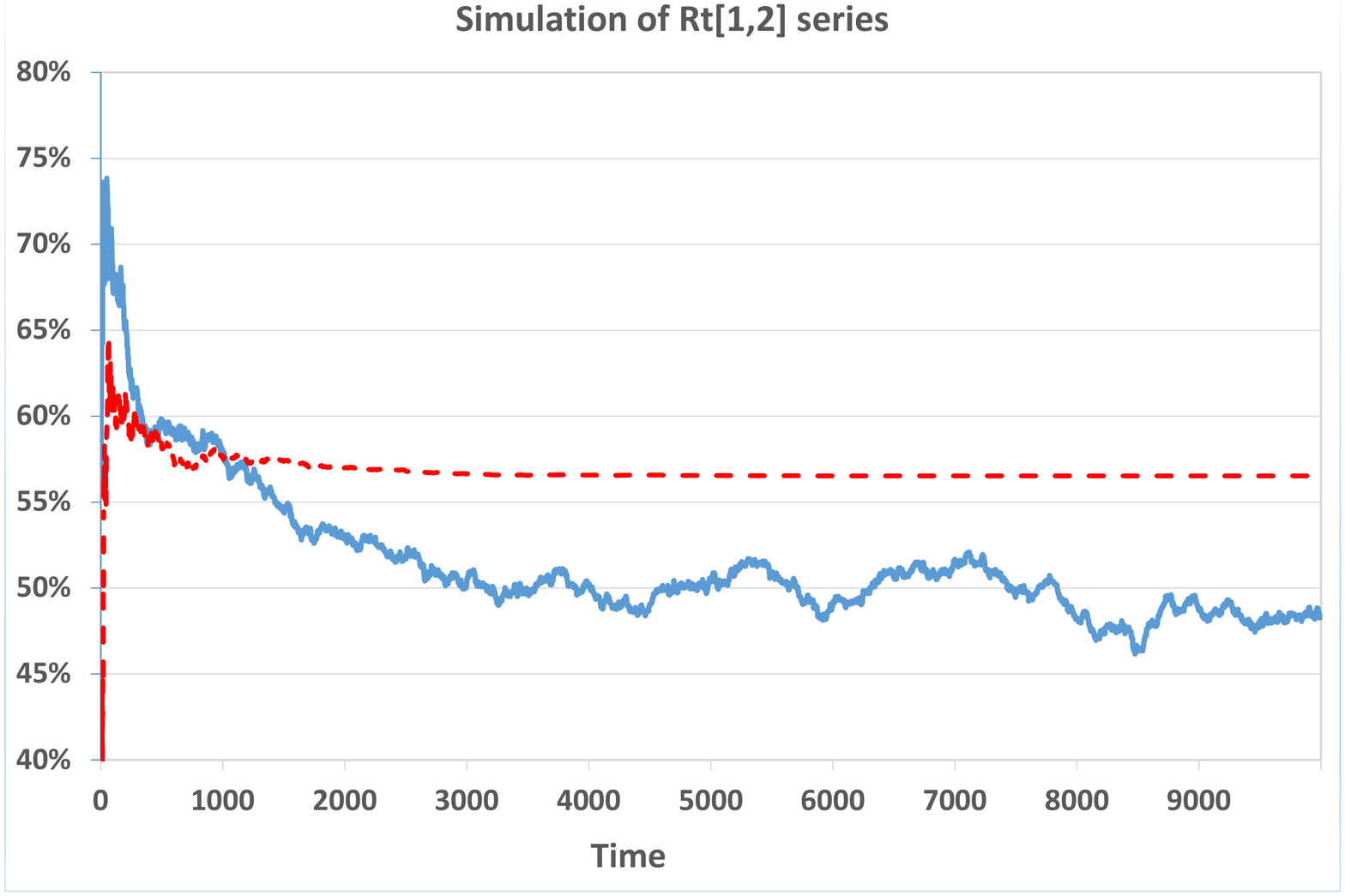}
\caption{Some simulated trajectories of $R_t[1,2]$ when $m^{(1)}=\sqrt{0.999}$ (solid line) or $m^{(1)}=\sqrt{1.001}$ (dashed line).} \label{RtFig}
\end{figure}

As expected, the value one is key for $m^{(1)}$. When the latter is very close to one but less than one ($(m^{(1)})^2=0.999$ in our case),
we check that the simulated trajectories of $(z_t)$, $(Q_t)$ and $(R_t)$ look stationary, once the influences of the starting values have been forgotten (broadly speaking when $t\geq 4000$): see Figure~\ref{QtFig} (solid lines).
On the other side, when this auto-regressive parameter is larger than one, even by a small amount ($(m^{(1)})^2=1.001$ in our case), we observe that the $(Q_t)$ trajectories explode: see Figure~\ref{QtFig} (dashed lines). Apparently, this is not the case for the $(R_t)$ correlation coefficients. They tend towards some constant levels (Figure~\ref{RtFig}), but differ from one experiment to another one. This phenomenon is a consequence of the normalization stage~(\ref{RtDCC}), but it seems difficult to maintain that feature will happen for almost every trajectory and any DCC model. Indeed, by managing increasingly (very) high numbers with the $(Q_t)$ process, we cannot exclude the possibility of spurious unexpected $(R_t)$ behaviors. Moreover, we have checked that the $(z_t)$ trajectories that we generated in this case do not seem to exhibit non-stationary patterns visually. This is logical because our DCC model tends towards a CCC model in such situations.
Therefore, it is likely that modelers will be able to manage (i.e. evaluate and simulate numerically) DCC trajectories in practice,
 even if~(\ref{cond2ThSimple}) or its generalizations are not satisfied. Actually, this task remains feasible as long as the numerical values of $(Q_t)$ are manageable (i.e. not too large) by our software. This is the case when the number of dates is not too large.

\mds

We have replaced Gaussian innovations $(\eta_t)$ by fat-tailed random vectors, to check to what extent this may be a source of instability.
The $\eta_{kt}$ components, $k=1,2$ and $t=0,\ldots,T$, have been drawn following mutually independent standardized Student laws with $\nu$ degrees of freedom, $\nu>0$.
We reiterate here that the $\eta_t$-moments of order $\nu$ or higher do not exist.
As long as $\nu \geq 2$ and $m^{(1)}<1$, we do not observe explosive patterns for $(z_t)$. The components of $R_t$ appear to become stationary, even if the
decrease in initial value effects is very slow when $\nu$ is close to two.
On the contrary, when $\nu< 2$, the processes $(Q_t)$ and $(R_t)$ are highly unstable. The former exhibit very spiky trajectories, while the latter often tend to be attracted
by the $1$ or $(-1)$ area.
Besides, some $(z_t)$ trajectories reach very high and unrealistic values ($10^{80}$ for instance).
In every case, when $m^{(1)}\geq 1$, the $(Q_t)$ trajectories explode and the $(R_t)$ ones tend to constant values that depend on each experiment. In such cases, the return process
$(z_t)$ may reach huge values, but only when $\nu \leq 2$, apparently. This analysis illustrates the necessity of considering innovations with finite second-order moments, and the fact that
higher-order moments are not mandatory to obtain stationary solutions.

\mds

Concerning the sufficient conditions that guarantee that stationary DCC solutions are unique, it is more difficult to evaluate their tightness
because they are more intricate and involve too many model characteristics. Nonetheless, with the simple scalar DCC model used in Example $2$, we observe that the
key condition~(\ref{Ex2Unicity}) will be more demanding when the number $m$ of underlyings increases. On the contrary, smaller values $|n^{(1)}|$ would help.
And the partially scalar case induces significantly less demanding conditions than the general case, because the values of the constants
$ C_\lambda^*$ and $C_q^*$ in the denominator are a lot higher than $C_\lambda$ and $C_q$ respectively. The effect of $|m^{(1)}|$ is ambiguous because it
appears in several quantities, especially $T_{33}$ and $C^*_\lambda$.

\mds

\section*{Acknowledgments}

\mds

{\rm We thank C. Francq and J.-M. Zako\"{\i}an for their valuable
remarks and discussions. Moreover, we are grateful to Eric Renault and two anonymous referees, who have proposed a number of ways of improving the article.
Finally, the authors thank the Labex ``Ecodec'' for its support.}

\appendix

\section{Technical lemmas}

We recall Tweedie's criterion, a key tool to prove the existence of
an invariant probability measure for a Markov chain. This result has
a remarkable advantage: contrary to more commonly used techniques
(based on some Lyapunov-Foster conditions, for example), it is not
necessary to state the irreducibility of the underlying Markov
chain, to obtain the existence of stationary solutions. Technically
speaking, proving the irreducibility of such a non-linear Markov
chain is a very challenging task in general.

\mds

Let $(X_t)_{t=1,2,\ldots}$ be a temporally homogeneous Markov chain
with a locally compact separable metric state space
$(S,\Bc)$. The transition probability is $P(x,A)=P(X_t \in A |\,
X_{t-1}=x)$, where $x\in S$ and $A\in \Bc$. Tweedie's
(1988) Theorem 2 provides:
\begin{lemma}
\label{TweedieLemma}

Suppose that $(X_t)$ is a Feller chain, i.e. for each bounded
continuous function $h$ on $S$, the function of $\xx$ given by
$E[h(X_{t}) \,|\, X_{t-1}=\xx]$ is also continuous.
\begin{enumerate}
\item If there exists, for some compact set $A\in \Bc$, a non-negative
function $g$ and $\eps>0$ satisfying
\begin{equation}
\label{TW1}
\int_{A^c} P(x,dy) g(y) \leq g(x) - \eps, \;\; x\in A^c,
\end{equation} then
there exists a $\sigma-$finite invariant measure $\mu$ for $P$
with $0<\mu (A) < \infty$.
\item Furthermore, if
\begin{equation}
\label{TW2}
\int_A \mu (dx) \left[ \int_{A^c} P(x,dy) g(y)   \right]<\infty,
\end{equation}
then $\mu$ is finite and hence $\pi=\mu/\mu(S)$ is an invariant
probability measure.
\item Furthermore, if
\begin{equation}
\label{TW3}
\int_{A^c} P(x,dy) g(y) \leq g(x) - f(x),\;\;x\in A^c,
\end{equation}
then $\mu$ admits a finite $f$-moment, that is
$ \int_S \mu(dy)f(y) <\infty.$
\end{enumerate}
\end{lemma}

\mds

The following Lemma is our version of Lemma A.2 in Ling and McAleer (2003).
Therefore, its proof is omitted.

\begin{lemma}
\label{LemmaM} For a given squared matrix $T$, if $\rho(|T|)<1$,
then there exists a vector $M>0$ such that $(Id - |T|')M
>0$.
\end{lemma}

\mds

\section{Proof of Proposition~\ref{prop_Markov}:}

Note that
$\eps_t$ (or $\vec{\eps}_t$, or even $Vech(\eps_t\eps_t')$) is a
function of the couple $(R_t,\eta_t)$ only. Due to~(\ref{RtDCC})
and~(\ref{Qtdef}), $R_t$ is a deterministic function of $X_{t-1}$.
Since $\eta_t$ is Markov with respect to $\Ic$, the law of $\eps_t$
knowing $\Ic_{t-1}$ is the law of $\eps_t$ knowing $X_{t-1}$ merely.
The same assertion applies with $T_t$, $\zeta_t$, or $X_t$ itself,
instead of $\eps_t$.

\mds

In other words, the non-linearity of the DCC model comes mainly
from $\vec{\eps}_t$ in $T_t$. But there exist constant matrices (of
zeros and ones) $F$ and $G$ such that~(\ref{defMarkov}) can be
rewritten
\begin{equation}
 X_t = (F \cdot \vec{\eps}_t^{\,*} ) \odot T_{o} X_{t-1}+(G \cdot Vech(\eps_t\eps_t')^* ) \odot \zeta_{o},
 \label{defMarkov2}
\end{equation}
where $T_{o}$ (resp. $\zeta_{o}$) is the $T_t$ matrix (resp.
$\zeta_t$ vector) when $\eps_t=1$, $\vec{\eps}_t^{\,*} := [\vec{\eps}_t^{\,'}, 1]'$ and
$Vech(\eps_t\eps_t')^* := [Vech(\eps_t\eps_t')',1]'$.
Since $\eps_t=R_t^{1/2}\eta_t$
and since $R_t$ is a measurable function of $X_{t-1}$, then $X_t$ is
clearly a function of $X_{t-1}$ and of the innovation $\eta_t$ only, that are Markov.
These arguments prove the Markovian structure of the $(X_t)$
process under A0.\; $\blacksquare$

\section{Proof of Theorem~\ref{thMoments}:}

First, let us check that $(X_t)$ is a Feller chain in a convenient space, to be able to apply Lemma~\ref{TweedieLemma} afterwards. Let $h$ be a bounded and
continuous function on $\RR^d$. Clearly,
\begin{eqnarray*}
\lefteqn{ E[h(X_{t})\,|\,X_{t-1}=\xx]= E[h(T_t\xx + \zeta_t)\,|\, X_{t-1}=\xx] }\\
&=& E[h(
\psi_1(\eps_t\eps_t')\xx + \psi_2(\eps_t\eps_t'))\,|\, X_{t-1}=\xx],
\end{eqnarray*}
for some continuous transforms $\psi_1$ and $\psi_2$. Note that
$\eps_t=R_t^{1/2}\eta_t$ and that $R_t^{1/2}$ is a continuous
function of $X_{t-1}$. Indeed, $R_t\mapsto R_t^{1/2}$ is continuous
(see Proposition 6.3 in Serre (2010), e.g.), and
$X_{t-1}\mapsto R_t$ is continuous by construction. Then,
$$ E[h(X_{t})\,|\,X_{t-1}=\xx]= E[h\circ \tilde\psi(\xx, \eta_t) \,| \, X_{t-1}=\xx]=
\int h\circ \tilde\psi(\xx, \eta) \, f_{\eta_t}(\eta | \xx) \, d\eta , $$
for some continuous transform $\tilde\psi$. Now, consider a sequence
of vectors $(\xx_n)$ that tends to $\xx$ when $n\rightarrow \infty$.
Since $h$ is bounded and since the sequence $(h\circ
\tilde\psi(\xx_n, \eta)f_{\eta_t}(\eta | \xx_n))_n$ is convergent for every $\eta$, we can
apply the dominated convergence theorem under E2. We deduce that $\xx \mapsto
E[h(X_{t})\,|\,X_{t-1}=\xx]$ is continuous and then $(X_t)$ is Feller.

\mds

Note that the vector $X_t$ belongs to the metric space $\RR^d$, endowed with the usual topology. Since we impose that the matrices $Q_t$ will be positive definite, $X_t$ will live in a subspace of $\RR^d$, where $X_t^{(3)}$ will gather only the components of definite positive matrices. It is easy to check that this subspace is separable and locally compact.
Therefore, the assumptions of Lemma~\ref{TweedieLemma} are satisfied.

\mds

Second, set $g(\xx)=1 + |\xx^{\otimes p}|' M$, for an arbitrary positive vector $M$, that will be chosen after.
Let us check that the latter function can be invoked as in Lemma~\ref{TweedieLemma}.
Clearly,
$$ E[g(X_t) \,|\, X_{t-1}=\xx]=1 + E\left[ | (T_t\xx +
\zeta_t)^{\otimes p}|'\,|\,X_{t-1}=\xx\right]M.$$ By expanding the Kronecker products, we can check that
$ (T_t\xx +\zeta_t)^{\otimes p} =  (T_t\xx)^{\otimes p}+ R(\xx),$
with
$$ \|R(\xx)\| \leq C_0\left( \|\zeta_t\|.\|(T_t\xx)^{\otimes
(p-1)}\|+\ldots + \|\zeta_t\|^{p-1}.\| (T_t\xx)\|+ \|\zeta_t\|^{p} \right),$$
for some positive constant $C_0$ and any multiplicative matrix norm $\|\cdot\|$.

\mds

Note that $(T_t\xx)^{\otimes k}=T_t^{\otimes k}.\xx^{\otimes k}$. Recall that $T_t$ is a function of $\vec{\eps}_t$, i.e. of $\eps_t$. Then, its conditional law depends on $R_{t}$, i.e. it is a function of $X_{t-1}$.
We deduce
\begin{eqnarray*}
\lefteqn{ E[ |(T_t\xx)^{\otimes p}| \,  |\, X_{t-1}=\xx]'M \leq
|\xx^{\otimes p}|'E[|T_t^{\otimes p}|'\,|\, X_{t-1}=\xx]M }\\
&\leq &
|\xx^{\otimes p}|'\left(\sup_{\xx \in \RR^d} E[|T_t^{\otimes p}|'\,|\, X_{t-1}=\xx] \right) M \\
&\leq & |\xx^{\otimes p}|'(T^*)' M.
\end{eqnarray*}
Now, choose $M$ as provided by Lemma~\ref{LemmaM}, when the matrix
$T$ in this lemma is replaced by $T^*$.

\mds

Moreover, $\eps_t=R_t^{1/2}\eta_t$, and the (positive definite)
matrix $R_t^{1/2}$ can be chosen so that all its coefficients are
less than $m^{1/2}$ (diagonalize this matrix on an orthonormal basis
and invoke Cauchy-Schwartz inequality). This implies that
constants $\alpha_k$ exist such that $\|Vech(\eps_t\eps_t')^{\otimes
k}\|\leq \alpha_k \|Vech(\eta_t\eta_t')^{\otimes k}\|$ when $k\leq
p$. Since $E[\|\eta_t\|^{2p} | X_{t-1}=\xx]<\bar h (\|\xx \|)$ by assumption, some
constants $c_{k,l}$ such that $E_{t-1}[\|\zeta_{t}\|^{k}.\|
\vec{\eps}_t \|^l]<c_{k,l} \bar h(\|\xx \|)^{(k+l)/p}$ for any couple $(k,l)$, $k+l\leq p$ exist. We
deduce the boundedness of $E[T_t^{\otimes k}\, | X_{t-1}=\xx]$, $k\leq p$, and
$$ E [\|R(\xx)\| \,|\, X_{t-1}=\xx] \leq C_1\left( \bar h (\|\xx \|)^{1/p} \|\xx^{\otimes (p-1)}\|+\ldots + \bar h (\|\xx \|)^{(p-1)/p}
\| \xx\| +\bar h (\|\xx \|)^p \right),$$
for some positive constant $C_1$.
Applying E2, we have obtained
\begin{eqnarray}
\lefteqn{E[g(X_t) \,|\, X_{t-1}=\xx] \leq 1 + |\xx^{\otimes p}|'(T^*)'M + O\left(\sum_{k=0}^{p-1} \| \xx^{\otimes k} \|\cdot \|\xx\|^{a} \right)     \nonumber }\\
&\leq & g(\xx) -|\xx^{\otimes p}|' \left( Id - (T^*)'\right) M + O\left(\sum_{k=0}^{p-1} \| \xx^{\otimes k} \|\cdot\|\xx\|^a \right),
\label{Et-1g}
\end{eqnarray}
for every constant $a>0$. By Lemma~\ref{LemmaM}, $( Id - (T^*)') M $ is strictly
positive. Then, a positive constant $c_0$ exists such that
$$ |\xx^{\otimes p}|' \left( Id - (T^*)'\right) M  \geq c_0 \sum_{j=1}^d |x_j|^p,$$ for every
$d$-dimensional vector $\xx$. Set $N(\xx):=\sum_{j=1}^d |x_j|^p$. By a
similar reasoning, a positive constant $c_1$ exists such that $
g(\xx) \geq c_1 N(\xx)$ for every $\xx\in \RR^d$. Moreover, by
applying H\"older's inequality, we have
$$\sum_{i_1,\ldots,i_k} |x_{i_1}\cdots x_{i_k}|=\left(\sum_{i=1}^d |x_i|\right)^k\leq
\left(\sum_{i=1}^d |x_i|^p\right)^{k/p} d^{k},$$
for every $k\leq p$.
Then, a positive constant $c_2$ exists such that
\begin{itemize}
\item $g(\xx) \leq1 + \| M \| \sum_{i_1,\ldots,i_p} |x_{i_1}\cdots
 x_{i_p}| \leq 1+ c_2 N(\xx)$, and
\item
every ``residual'' term $\| \xx^{\otimes k} \|$ is bounded above by (a scalar times)
$N(\xx)^{k/p}$, when $k< p$.
\end{itemize}
Therefore, this provides
\begin{eqnarray*}
\lefteqn{E[g(X_t) \,|\, X_{t-1}=\xx]
\leq  g(\xx) \left[ 1 -c_0 \frac{N(\xx)}{g(\xx)} + O\left( \sup_{k=0,\ldots,p-1} \frac{N(\xx)^{(k+a)/p}}{g(\xx)} \right)\right] }\\
&\leq &  g(\xx) \left[ 1 - \frac{c_0 N(\xx)}{1+c_2 N(\xx)} + O\left( \sup_{k=0,\ldots,p-1} \frac{N(\xx)^{(k+a)/p}}{c_1 N(\xx)} \right)\right].
\end{eqnarray*}
Let us define the set $A:=\{ \xx \in \RR^d \,|\, N(\xx) \leq \Delta
\}$, for some $\Delta >1$. When $\Delta$ is sufficiently large, we
obtain, for any $\xx\not\in A$ and a power $a$ s.t. $0<a< 1/p$,
\begin{equation}
0\leq E[g(X_t) \,|\, X_{t-1}=\xx]
\leq   g(\xx) \left[ 1 - \frac{c_0}{2c_2} + O\left( \frac{\Delta^{a-1/p}}{c_1 } \right)\right]  < g(\xx) \left[ 1 - \frac{c_0}{3c_2} \right].
\label{Et-1gg}
\end{equation}
Since $g(\xx)\geq 1$, it follows that $E[g(X_t  ) \,|\, X_{t-1}=\xx] \leq g(\xx) -\eps$ for some $\eps>0$.
This proves Equation~(\ref{TW1}) in
Lemma~\ref{TweedieLemma}. Therefore, there exists a $\sigma$-finite invariant measure $\mu$ for the Markov chain $(X_t)$, and $0<\mu(A)<\infty$.

\mds

For any $\xx\in A$, Equation~(\ref{Et-1g}) provides
$$E[g(X_t) \,|\, X_{t-1}=\xx]
\leq  g(\xx) + O\left(\sum_{k=0}^{p-1} \| \xx^{\otimes k} \|\cdot\|\xx\|^a \right)\leq c_3\Delta^{(1+a/p)} $$
for some constant $c_3$ that does not depend on $\xx$. Then,
\begin{equation*}
\int_A \mu (dx) \left[ \int_{A^c} P(x,dy) g(y)   \right]\leq \int_A \mu (dx) E[g(X_t) \,|\, X_{t-1}=\xx] \leq C \Delta^{1+a/p} \mu(A) <\infty.
\end{equation*}
We deduce that $\mu$ is finite and hence $\pi=\mu/\mu(\RR^d)$ is an invariant
probability measure of $(X_t)$. This implies that a strictly stationary solution satisfying~(\ref{defMarkov}) exists, still denoted by $(X_t)$.

\mds

Third, by invoking Equation~(\ref{Et-1gg}), we get~(\ref{TW3}) in Lemma~\ref{TweedieLemma} with $f(\xx)=\beta g(\xx)$, for some $\beta \in (0,1)$.
Since $g(\xx)\geq c_1 N(\xx)$, we obtain
\begin{equation}
 E_\pi[ N(X_{t})]<\infty.
\label{MomentsX}
\end{equation}
In particular, invoking H\"older's inequality, this implies that $
E_\pi[ z_{it}^{2k}]<\infty$, for every $i=1,\ldots,m$ and every
$k\leq p$. $\blacksquare$

\mds

\begin{remark}
Equation~(\ref{MomentsX}) provides a lot more than only the finiteness of $z$'s moments. Overall, it means that
$$ E_\pi\left[ \sum_{i=1}^m h_{it}^{p} \right]<\infty,\;\; E_\pi\left[ \sum_{i=1}^m z_{it}^{2p} \right]<\infty, $$
$$ E_\pi\left[ \sum_{i,j=1}^{m} |Q_{ij,t}|^{p} \right]<\infty,\;\text{and}\; E_\pi\left[ \sum_{i=1}^{m} |\eps_{it}|^{2p} \right]<\infty. $$
\end{remark}

\mds

\section{Proof of Theorem~\ref{thSimple}:}

Let us consider $\lambda$, a non zero eigenvalue of $T^*$, when $p=1$. We can easily check that this matrix is simply $T_t$, by replacing $\vec{\eps}_t$ by one, and replacing the coefficients of the matrices $\tilde M_k$ and $\tilde N_l$ by their absolute values
(remember that the matrices $A_i$ and $B_j$ are already non-negative). Let $\vv=(\vv^{(1)},\vv^{(2)},\vv^{(3)},\vv^{(4)})$ be the associated eigenvector, where the dimensions of the subvectors $\vv^{(k)}$, $k=1,\ldots,4$ are consistent with those of
$X_t$ in~(\ref{defXt}). We can further split the latter subvectors, so that they comply with the matrices $A_i$, $B_j$, $\tilde M_k$ and $\tilde N_l$. With obvious vector sizes, we will denote $\vv^{(1)} = (\vv^{(1)}_1,\ldots,\vv^{(1)}_r)$, $\vv^{(2)} = (\vv^{(2)}_1,\ldots,\vv^{(2)}_s)$, $\vv^{(3)} = (\vv^{(3)}_1,\ldots,\vv^{(3)}_\nu)$ and $\vv^{(4)} = (\vv^{(4)}_1,\ldots,\vv^{(4)}_\mu)$.

\mds

By simple block-matrix calculations, the relation $T^* \vv = \lambda \vv$ implies
$$ \vv_1^{(1)} = \vv_1^{(2)} = \sum_{i=1}^r \frac{A_i \vv^{(1)}_1}{\lambda^i}  +
 \sum_{j=1}^s  \frac{B_j\vv^{(2)}_1}{\lambda^j},$$
 $$
 \vv_1^{(3)} = \sum_{k=1}^\nu  \frac{\tilde M_k\vv^{(3)}_1}{\lambda^k}  +
 \sum_{l=1}^\mu  \frac{\tilde N_l \vv^{(4)}_1}{\lambda^l},\;\text{and}\;\;
 \vv_1^{(4)} = 0.$$
Note that $\vv^{(1)}_i =\vv^{(1)}_1/ \lambda^i$ and $\vv^{(2)}_j =\vv^{(2)}_1/ \lambda^j$ for every $i$ and $j$.
Moreover, $\vv^{(3)}_k =\vv^{(3)}_1/ \lambda^k$ and $\vv^{(4)}_l =0$ for every $k$ and $l$.
\mds

If $\lambda \geq 1$, then
\begin{eqnarray*}
\lefteqn{ \Nc( \vv_1^{(1)} ) \leq  \sum_{i=1}^r \| A_i \|_{\Nc} \frac{\Nc(\vv^{(1)}_1)}{|\lambda|^i}  +
 \sum_{j=1}^s \|B_j\|_{\Nc} \frac{\Nc(\vv^{(1)}_1)}{|\lambda|^j}  }\\
&
\leq & \Nc(\vv^{(1)}_1)\left( \sum_{i=1}^r \| A_i \|_{\Nc}   +
 \sum_{j=1}^s \|B_j\|_{\Nc} \right).
\end{eqnarray*}
Similarly,
$$ \Nc^*( \vv_1^{(3)} )
\leq  \sum_{k=1}^\nu \| \tilde M_k \|_{\Nc^*} \frac{\Nc^*(\vv^{(3)}_1)}{|\lambda|^k}
\leq \Nc^*( \vv^{(3)}_1 ) \sum_{k=1}^\nu \| \tilde M_k  \|_{\Nc^*}.$$
Since $\vv\neq 0$, we obtain
$$ 1 \leq
\sum_{i=1}^r \| A_i \|_{\Nc}   +
 \sum_{j=1}^s \|B_j\|_{\Nc}, \;\text{or} \;
1 \leq  \sum_{k=1}^\nu \| \tilde M_k \|_{\Nc^*}.$$
This proves the result. $\blacksquare$

\mds

\section{Proof of Theorem~\ref{ThUnicity}:}

Suppose that two strongly stationary solutions $(X_t)$ and
$(\tilde{X}_t)$ exist. Since both satisfy
Equation~(\ref{defMarkov}), with obvious notations, we can write for
every $t$
$$  X_t = T_t. X_{t-1}+\zeta_t,\;\text{and}\;  \tilde X_t = \tilde T_t. \tilde X_{t-1}+\tilde\zeta_t.$$
Note that the difference between $T_t$ and $\tilde T_t$ is only due to the (a priori different) factors $\eps_t$ and $\tilde\eps_t$.
We want to prove that, for every $t$, almost certainly $X_t=\tilde{X}_t$.

\mds

The problem will be solved if we prove the uniqueness of the process $(X_t^{(3)},X_t^{(4)})$, given by subvectors of $(X_t)$. For the moment, assume it has been proved. Recall that
$$ X_t^{(3)}:= (Vech(Q_t),\ldots, Vech(Q_{t-\nu+1}))',\;\text{and}$$
$$ X_t^{(4)} := (Vech(\eps_t\eps_t'),\ldots,Vech(\eps_{t-\mu+1}\eps_{t-\mu+1}'))'.$$
$(R_t)$ is therefore unique, due to~(\ref{RtDCC}). Moreover, the sequence
of random matrices $(T_t)$ and of noises $(\zeta_t)$ are
also unique, similarly to the CCC case. Now, let us prove the uniqueness
of $Y_t:=(X_t^{(1)},X_t^{(2)})$, knowing $(\eta_t)$. This would
imply the uniqueness of the instantaneous volatility process $(D_t)$
and of the return process $(z_t)$ themselves. With our notations, we
have
$$ Y_t = \bar T_t Y_{t-1} + \bar \zeta_t, \; \text{and}\; \tilde{Y}_t = \bar T_t \tilde Y_{t-1} + \bar \zeta_t,$$
for every $t$, by setting $\bar\zeta_t =
(\zeta_t^{(1)},\zeta_t^{(2)}).$ The arguments are then standard: for
instance, see Theorem 2.4's proof in Francq and Zako\"{\i}an (2010).

\mds

We recall the reasoning briefly, to get the uniqueness of $(Y_t)$. Note that
$$ Y_t - \tilde{Y}_t = \bar T_t \bar T_{t-1}\cdots \bar T_{t-p} \cdot (Y_{t-p-1} - \tilde Y_{t-p-1}),$$
whenever $p>1$. Since the sequences $(Y_t)$ and $(\tilde Y_t)$ are stationary,
it is sufficient to prove that $\| \bar T_t \bar T_{t-1}\cdots \bar T_{t-p} \|$ tends
to zero a.e. when $p$ tends to the infinity, for any matrix norm.
This is the case under Assumption U3 because, for every sequence
$(\eps_t)$,
$$ E[\ln \|\bar T_t \bar T_{t-1}\ldots \bar T_1 \|_{1} ]
\leq
\ln E[ \|\bar T_t \bar T_{t-1}\ldots \bar T_1 \|_{1} ]=
\ln  \|\left(  \bar T_t^* \right)^t \|_{1}, $$
by invoking Jensen's inequality, the stationarity of $(\bar T_t)$, and by noting that all the coefficients of the matrices $\bar T_t$ are non-negative.
It is well-known that $\lim_{t\rightarrow \infty} t^{-1} \ln ( \|A^t \|) =  \ln \rho( A),$ for any squared matrix $A$.
Apply this result with $A=  \bar T_t^*$.
Therefore $\gamma_T = \lim_{t\rightarrow
\infty} t^{-1} E[\ln \|\bar T_t \bar T_{t-1}\ldots \bar T_1 \|_{1}] $, the top Lyapunov exponent of
the sequence of random matrices $(\bar T_t)$, is strictly negative under Assumption U3.
Since the sequence of matrices $(\bar T_t)$ is strictly stationary under U0, we get
$\lim_{p \rightarrow +\infty} \| \bar T_t \bar T_{t-1}\cdots \bar T_{t-p} \|_1 = 0$ with probability one (Theorem 2.3 in Francq and Zako\"{\i}an, 2010).
This provides the
uniqueness of the processes $(D_t)$ and $(z_t)$, once we assume the uniqueness of the processes $(Q_t)$ and $(\eps_t)$.

\mds

Now, let us prove the uniqueness of $(X_t^{(3)},X_t^{(4)})$ or, in
other terms, of $(Q_t,\eps_t)$. This task is clearly more tricky, because
we will have to deal with the nonlinear feature of the DCC
specification. Here, the convenient matrix norm will be the spectral
norm $\|\cdot \|_s$. Consider two stationary solutions
$(Q_t,\eps_t)$ and $(\tilde Q_t,\tilde \eps_t)$. Since the spectral
norm is sub-multiplicative, we deduce from~(\ref{Qtdef}) that
\begin{eqnarray}
\lefteqn{ \| Q_t - \tilde Q_t \|_s \leq \sum_{k=1}^\nu \| M_k \|_s^2  \| Q_{t-k} - \tilde Q_{t-k}\|_s \nonumber}\\
&+&
\sum_{l=1}^\mu \| N_l \|_s^2 \|\eps_{t-l} \eps_{t-l}' - \tilde\eps_{t-l} \tilde\eps_{t-l}' \|_s.
\label{QtIneg}
\end{eqnarray}
The key point will be to bound from above the terms $ \|\eps_{t-l}
\eps_{t-l}' - \tilde\eps_{t-l} \tilde\eps_{t-l}' \|_s$ by a function
of $ \| Q_{t-l} - \tilde Q_{t-l}\|_s$. To lighten the indices, we
assume $l=0$. Clearly, we have
\begin{eqnarray*}
\lefteqn{ \|\eps_{t} \eps_{t}' - \tilde\eps_{t} \tilde\eps_{t}' \|_s =
\|R_t^{1/2}\eta_{t} \eta_{t}' R_t^{1/2}- \tilde R_t^{1/2}\eta_{t} \eta_{t}' \tilde R_t^{1/2}\|_s   }\\
&\leq &
\|(R_t^{1/2}-\tilde R_t^{1/2}) \eta_{t} \eta_{t}' R_t^{1/2} \|_s
+ \| \tilde R_t^{1/2}\eta_{t} \eta_{t}' (R_t^{1/2} -\tilde R_t^{1/2})\|_s   \\
&\leq &
\|R_t^{1/2}-\tilde R_t^{1/2} \|_s  \| \eta_{t} \eta_{t}'\|_s  \|R_t^{1/2} \|_s
+ \| \tilde R_t^{1/2}\|_s  \|\eta_{t} \eta_{t}' \|_s  \|R_t^{1/2} -\tilde R_t^{1/2}\|_s.
\end{eqnarray*}
Since the rank of $\eta_t\eta_t'$ is one,
$\|\eta_t\eta_t'\|_s=Tr(\eta_t\eta_t')=\|\eta_t \|_2^2$.
Moreover,
$$\| R_t^{1/2} \|_s=\rho(R_t)^{1/2}\leq Tr(R_t)^{1/2}=\sqrt{m}.$$
We deduce
\begin{equation}
\|\eps_{t} \eps_{t}' - \tilde\eps_{t} \tilde\eps_{t}' \|_s \leq 2m^{1/2} \| \eta_t \|_2^2. \|R_t^{1/2}-\tilde R_t^{1/2} \|_s.
\label{InegEpsEps}
\end{equation}
Since the spectral norm is unitarily invariant, Theorem 6.2 in Hingham (2008) provides
\begin{equation}
\|R_t^{1/2}-\tilde R_t^{1/2} \|_s \leq \frac{1}{\lambda_{1}(R_t)^{1/2} + \lambda_{1}(\tilde R_t)^{1/2} }
\|R_t-\tilde R_t \|_s.
\label{RootRtIneg}
\end{equation}

Note that, for any $t$,
\begin{eqnarray*}
\lefteqn{ \lambda_1(R_t) = \min_\xx \frac{\xx' R_t \xx}{\xx' \xx} = \min_\xx \frac{\xx' diag(Q_t)^{-1/2} Q_t diag(Q_t)^{-1/2} \xx}{\xx' \xx} }\\
&\geq &\min_\yy \frac{\yy'  Q_t  \yy}{\yy' \yy}
\min_\xx \frac{\| diag(Q_t)^{-1/2} \xx \|^2_2 }{\|\xx \|_2^2} \\
&\geq & \lambda_1(Q_t) \min_i \frac{1}{q_{ii,t}} \geq \frac{ C_\lambda }{ \| Q_t \|_{\max}},
\end{eqnarray*}
invoking Lemma~\ref{LambdaBounded}. Since the same inequality applies with $\lambda_1(\tilde R_t)$, we get
\begin{equation}
\frac{1}{\lambda_{1}(R_t)^{1/2} + \lambda_{1}(\tilde R_t)^{1/2} }\leq
\frac{\| Q_t \|^{1/2}_{\max} + \| \tilde Q_t \|^{1/2}_{\max} }{\sqrt{C_\lambda}}\cdot
\label{BoundLambda2}
\end{equation}
Moreover,
\begin{eqnarray*}
\lefteqn{ R_t-\tilde R_t = (diag(Q_t)^{-1/2}- diag(\tilde Q_t)^{-1/2}) Q_t diag(Q_t)^{-1/2} }\\
&+&
diag(\tilde Q_t)^{-1/2}(Q_t-\tilde Q_t)diag(Q_t)^{-1/2} \\
&+&
 diag(\tilde Q_t)^{-1/2}  \tilde Q_t (diag(Q_t)^{-1/2}- diag(\tilde Q_t)^{-1/2}) := \Rc_1 + \Rc_2 + \Rc_3.
\end{eqnarray*}

Note that $ \Rc_1 = [  (q_{ii,t} - \tilde q_{ii,t})q_{ij,t}
q^{-1/2}_{jj,t}  q^{-1/2}_{ii,t} \tilde q^{-1/2}_{ii,t}
/(q^{1/2}_{ii,t} + \tilde q^{1/2}_{ii,t})]_{1\leq i,j\leq m}$ and
$|q_{ij,t}|\leq \sqrt{ q_{ii,t}}\sqrt{q_{jj,t}}$ (Cauchy-Schwartz).
Since $\|A\|_{\max} \leq \|A\|_s \leq m \|A\|_{\max} $, we get
$$\|\Rc_1\|_{\max} \leq C_q^{-1}\| diag(q_{ii,t} - \tilde q_{ii,t})   \|_{\max} \leq C_q^{-1}\| Q_t - \tilde Q_t \|_{\max}
\leq C_q^{-1}\| Q_t - \tilde Q_t \|_s ,$$ and $\|\Rc_1\|_s \leq m C_q^{-1}\| Q_t - \tilde Q_t \|_s $.
Similarly, $\|\Rc_3\|_s \leq m C_q^{-1}\| Q_t - \tilde Q_t \|_s$.
By Lemma~\ref{LambdaBounded}, we obtain
\begin{eqnarray*}
\lefteqn{ \|\Rc_2\|_s = \| diag(\tilde Q_t)^{-1/2}(Q_t-\tilde Q_t)diag(Q_t)^{-1/2} \|_s }\\
& \leq &
\| diag(\tilde Q_t)^{-1/2} \|_s \| diag( Q_t)^{-1/2} \|_s  \| Q_t-\tilde Q_t \|_s \\
&\leq & \frac{1}{\sqrt{\min_i q_{ii,t}}} \frac{1}{\sqrt{\min_i \tilde q_{ii,t}}} \| Q_t - \tilde Q_t \|_s
\leq \frac{1}{C_q} \| Q_t - \tilde Q_t \|_s. \hspace{5cm}
\end{eqnarray*}

Globally, we get
\begin{equation}
\| R_t-\tilde R_t \|_s \leq \frac{2m+1}{C_q} \| Q_t - \tilde Q_t \|_s
\label{RtGlobal}
\end{equation}
everywhere. Recalling~(\ref{InegEpsEps}),~(\ref{RootRtIneg})~(\ref{BoundLambda2}) and~(\ref{RtGlobal}), we deduce
\begin{equation}
\|\eps_{t} \eps_{t}' - \tilde\eps_{t} \tilde\eps_{t}' \|_s \leq
\frac{2m^{1/2} \|\eta_t \|_2^2}{\sqrt{C_\lambda}} \cdot \frac{2m+1}{C_q} (\|Q_t\|^{1/2}_{\max} + \| \tilde Q_t \|^{1/2}_{\max})  \| Q_t - \tilde Q_t \|_s.
\end{equation}

\mds

Set $v_t:=\| Q_t - \tilde Q_t \|_s$. By using the previous inequality and the notation of Lemma~\ref{QtBounded}, we obtain
\begin{equation}
v_t \leq \sum_{k=1}^\nu \| M_k \|_s^2 v_{t-k}
+\sum_{l=1}^\mu \| N_l \|_s^2
\frac{4m^{1/2}(2m+1) }{\sqrt{C_\lambda}C_q}
\|\eta_{t-l} \|_2^2 \sqrt{q_t} v_{t-l}:=\sum_{j=1}^{\kappa} \beta_{j,t} v_{t-j},
\label{QtInegFinal}
\end{equation}
for all $t$ and with our notations.

\mds

Setting $\vec{v}_t := [v_t,v_{t-1},\ldots,v_{t-\kappa +1}]'$, we get
$$ 0\leq \vec{v}_t \leq N^*_t \vec{v}_{t-1} \leq \ldots \leq N_t^* N_{t-1}^* \cdots N_{t-p}^* \vec{v}_{t-p-1},$$
for any positive integer $p$. By the stationarity of the $(Q_t)$ and $(\tilde Q_t)$ trajectories,
the norm of $\vec{v}_t$ is bounded by a constant that is independent of $t$.
Moreover, under the assumptions U0 and U2,
$\| N_t^* N_{t-1}^*\cdots  N_{t-p}^*\|_s$ tends to zero a.e.
when $p\rightarrow +\infty$ and for any fixed $t$ (see Francq and Zako\"{\i}an 2010, Theorem 2.3). We deduce $v_t\rightarrow 0$ a.e. when
$t\rightarrow \infty$,
because $(v_t)$ can be initialized arbitrarily far in the past. This implies that $Q_t = \tilde Q_t$ a.e. Therefore, $R_t = \tilde R_t$ a.e. and $\eps_t = \tilde\eps_t$
a.e., knowing $(\eta_t)$. This concludes the proof of uniqueness. The ergodicity of the (now unique) DCC solution is a consequence of Corollary 7.17 in Douc et al. (2014). $\blacksquare$

\mds

\begin{lemma}
\label{QtBounded} Under Assumption U0-U1, for almost every trajectory
of a solution $(Q_t)$ of the DCC model, we have
$$ \|Q_t\|_{\max} \leq
\frac{\| Vech(W_0) \|_2}{1- \| T_{33}\|_s} +  \sqrt{\frac{m^3(m+1)}{2}} \sum_{l=1}^\mu \| \tilde N_l\|_{s}\xi_{t-l}:=q_t,$$
where
$\xi_t := \sum_{k=0}^{+\infty} \| T_{33}\|_s^k  \|\eta_{t-k}\|_2^2.$
\end{lemma}

\mds

If these innovations $|\eta_t|$
are bounded from above by a positive constant $C_\eta$ a.e., then
the latter inequality is simply
$$ \|Q_t\|_{\max} \leq
\frac{\| Vech(W_0) \|_2 +  \sqrt{\frac{m^3(m+1)}{2}} \sum_{l=1}^\mu \| \tilde N_l\|_{s} C_\eta^2   }{1- \| T_{33}\|_s}.$$

\mds

{\bf Proof of Lemma~\ref{QtBounded}:}
For any $t$, $ \|Vech(\eps_{t} \eps_{t}')\|_s = \|Vech(\eps_{t}
\eps_{t}')\|_2 \leq \sqrt{m(m+1)} \| \eps_t\|_\infty^2/ \sqrt{2}.$
Moreover, since
$\|\xx\|_s=\|\xx\|_2$ for any vector $\xx$ and $\|A\|_{\max}\leq
\| A\|_s$ for any matrix $A$ (L\"utkepohl, 1996, p. 111), we get
\begin{eqnarray*}
\lefteqn{\| \eps_t\|_\infty  \leq \| \eps_t\|_s  \leq \| R_t^{1/2} \eta_t\|_s
 \leq \| R_t^{1/2} \|_s. \|\eta_t\|_s }\\
 &\leq & \| R_t\|_s^{1/2} \| \eta_t\|_2  \leq   \sqrt{m} \| \eta_t\|_2 .
\end{eqnarray*}
This proves the inequality
$ \|Vech(\eps_{t-l} \eps_{t-l}')\|_s \leq \sqrt{m^3(m+1)}
 \|\eta_{t-l} \|_2^2/\sqrt{2},$ for every $t$ and $l$.

\mds

With the notations of Subsection~\ref{DCCMarkov}, consider the dynamics of the random vector
$ X_t^{(3)}:= (Vech(Q_t),\ldots, Vech(Q_{t-\nu+1}))'.$ Clearly,
$ X_t^{(3)} = T_{33} X_{t-1}^{(3)}+ \pi_t,$
where $$ \pi_t:= Vech(W_0)+\sum_{l=1}^\mu \tilde N_l Vech(\eps_{t-l}
\eps_{t-l}').$$

\mds

We deduce from U1 that
\begin{eqnarray*}
\lefteqn{ \|Q_t\|_{\max} \leq \|X_{t}^{(3)}\|_{\max} \leq \|X_{t}^{(3)}\|_s \leq
 \sum_{k=0}^{+\infty} \| T_{33}\|_s^k  \| \pi_{t-k} \|_s   }\\
& \leq & \sum_{k=0}^{+\infty} \| T_{33}\|_s^k     \left\{
\| Vech(W_0) \|_s +\sum_{l=1}^\mu \| \tilde N_l\|_{s} .\| Vech(\eps_{t-k-l} \eps_{t-l}')\|_s \right\} \\
&\leq &
\frac{\| Vech(W_0) \|_s}{1- \| T_{33}\|_s} + \sum_{k=0}^{+\infty} \| T_{33}\|_s^k \sum_{l=1}^\mu \| \tilde N_l\|_{s} . \sqrt{\frac{m^3(m+1)}{2}}  \|\eta_{t-k-l} \|_2^2 \\
&\leq &
\frac{\| Vech(W_0) \|_s}{1- \| T_{33}\|_s} +  \sqrt{\frac{m^3(m+1)}{2}} \sum_{l=1}^\mu \| \tilde N_l\|_{s}\xi_{t-l}:= q_t.
\end{eqnarray*}
Since the spectral norm of $Vech(W_0)$ is its Euclidian norm, as for any vector, we obtain the result.\; $\blacksquare$

\mds

\begin{lemma}
\label{LambdaBounded} Under Assumption U1, for almost every
trajectory of a solution $(Q_t)$ of the DCC model, we have
$\lambda_1(Q_t) \geq C_\lambda$ and $ \min_{i=1,\ldots,m} q_{ii,t}\geq C_q$,
where $C_\lambda =\lambda_1(W_0)$ and
$C_q:=\min_{i=1,\ldots,m}(W_0)_{ii}$. In addition, if we assume U4,
then
$\lambda_1(Q_t) \geq C^*_\lambda$ and $ \min_{i=1,\ldots,m} q_{ii,t}\geq C^*_q$, with
$$C_\lambda^* := \frac{ \lambda_1(W_0)}{1- \sum_{k=1}^\nu (m^{(k)})^2} \;\; \text{and}
\;\;
C_q^* := \frac{\min_{i=1,\ldots,m}(W_0)_{ii}}{1- \sum_{k=1}^\nu (m^{(k)})^2}\cdot$$
\end{lemma}

\mds

{\bf Proof of Lemma~\ref{LambdaBounded}:} Is it known that, for any
two positive definite matrices $A$ and $B$,
$\lambda_1(A+B)\geq \lambda_1(A) + \lambda_1(B)$ (Weyl's Theorem.
See L\"utkepohl, 1996, p. 75). In our case, we deduce that
$\lambda_1(Q_t)\geq \lambda_1(W_0)$ everywhere, due to
Equation~(\ref{Qtdef}).

\mds

We can improve this lower bound in the particular case of
``partially'' scalar DCC models. Indeed, in this case, we have
\begin{equation}
  \lambda_1(Q_t)\geq \lambda_1(W_0) + \sum_{k=1}^\nu \lambda_1 ((m^{(k)})^2 Q_{t-k}) \geq   \lambda_1(W_0) +
  \sum_{k=1}^\nu (m^{(k)})^2 \lambda_1 ( Q_{t-k}).
\label{InegLambda}
\end{equation}
Introduce the random vector $\vec{\lambda}_t :=
(\lambda_1(Q_t),\ldots,\lambda_1(Q_{t-\nu+1}))'$ and
$\vec{\lambda}_W:=(\lambda_1(W_0),0,\ldots,0)'$. Because
of~(\ref{InegLambda}), we have
$\vec{\lambda}_t  \geq M^* \vec{\lambda}_{t-1} + \vec{\lambda}_W$ for every $t$.
Under Assumption U4, it is easy to check that $\sum_{k=0}^{+\infty}
(M^*)^k $ is absolutely convergent and that
$$\vec{\lambda}_t  \geq \sum_{k=0}^{+\infty}
(M^*)^k \vec{\lambda}_W:=\vec{\lambda}_\infty,$$ for every $t$.
Obviously, $M^* \vec{\lambda}_\infty + \vec{\lambda}_W =
\vec{\lambda}_\infty.$ Due to the definition of $M^*$, this implies that all the components of
$\vec{\lambda}_\infty$ are the same, i.e. a real number
$\lambda_\infty$ exists such that $\vec{\lambda}_\infty = \lambda_\infty
e$, $e\in \RR^\nu$. Taking the first component of the vectorial
equation $\lambda_\infty M^* e + \vec{\lambda}_W = \lambda_\infty e$
provides $\lambda_\infty \sum_{k=1}^\nu (m^{(k)})^2 +
\lambda_1(W_0)= \lambda_\infty$. This states the lower bound of
$\lambda_1(Q_t)$ under U4.

\mds

Consider a fixed index $i=1,\ldots,m$. The reasoning for the sequence $(q_{ii,t})_t$ is similar, because
$$  q_{ii,t}\geq (W_0)_{ii} + \sum_{k=1}^\nu (m^{(k)})^2  q_{ii,t-k},$$
for all $t$, as this inequality is playing the same role
as~(\ref{InegLambda}). This implies the desired result.
$\blacksquare$

\mds

\bigskip

\noindent {\large {\it REFERENCES}}

\bigskip

\noindent  Aielli, G.P. (2013). Dynamic Conditional Correlation: on
properties and estimation. {\it Journal of Business \& Economic
Statistics} 31, 282-299.

\noindent Billio, M. \& Caporin, M. (2005). Multivariate Markov
switching dynamic conditional correlation GARCH representations for
contagion analysis. {\it Statistical Methods \& Applications} 14,
145-161.

\noindent Billio, M., M. Caporin \& M. Gobbo (2006). Flexible
dynamic conditional correlation multivariate GARCH for asset
allocation. {\it Applied Financial Economics Letters} 2, 123-130.

\noindent Bollerslev, T. (1990). Modeling the Coherence in Short Run
Nominal Exchange Rates: A Multivariate generalized ARCH Model. {\it
The Review of Economics and Statistics} {\rm 72}, 498-505.

\noindent Bougerol, P. \& Picard, N. (1992). Stationarity of GARCH processes and of some nonnegative time series. {\it
Journal of Econometrics} 52, 1714-1729.

\noindent Boussama, F., F. Fuchs \& R. Stelzer (2011). Stationarity
and Geometric Ergodicity of BEKK Multivariate GARCH Models. {\it
Stochastic Processes and their Applications} {\rm 121}, 2331-2360.

\noindent Cappiello L., R.F. Engle \& K. Sheppard (2006). Asymmetric
dynamics in the correlations of global equity and bond returns. {\it
Journal of Financial Econometrics} 4, 537-572.

\noindent Caporin, M. \& M. McAleer (2013). Ten Things You Should
Know about the Dynamic Conditional Correlation Representation. {\it
Econometrics} {\rm 1}, 115-126.

\noindent
Comte, F. \& O. Lieberman (2003).
Asymptotic theory for multivariate GARCH processes. {\it Journal of Multivariate
Analysis} 84, 61-84.

\noindent
Donald, S.G., Imbens, G.W. \& W.K. Newey (2003).
Empirical likelihood estimation and consistent tests with conditional moment restrictions. {\it Journal of Econometrics} 117, 55-93.

\noindent Douc, R., E. Moulines \& D. Stoffer (2014). {\it Nonlinear Time Series}. Chapman \& Hall.

\noindent Engle, R.F. \& F.K. Kroner (1995). Multivariate
simultaneous generalized ARCH. \textit{Econometric Theory} {\rm 11},
122-150.

\noindent Engle, R.F. \& K. Sheppard (2001). Theoretical and
empirical properties of dynamic conditional correlation multivariate
GARCH. Working Paper 2001-15, University of California at San Diego.

\noindent Engle, R.F. (2002). Dynamic conditional correlation: a
simple class of multivariate GARCH models. \textit{Journal of
Business and Economic Statistics} {\rm 20}, 339-350.

\noindent Fermanian, J.-D. \& H. Malongo (2013). On the link between
instantaneous volatilities, switching regime probabilities and
correlation dynamics. Working paper Crest.

\noindent Francq, C. \& J.-M. Zako\"{\i}an (2010). {\it GARCH
models}. Wiley.

\noindent Franses, P.H. \& C.M. Hafner (2009). A generalized dynamic
conditional correlation model: Simulation and application to many
assets. {\it Econometric Reviews} 28, 612-631.

\noindent Higham, N.J. (2008). {\it Functions of Matrices}. SIAM.

\noindent Kasch, M. \& M. Caporin (2013). Volatility threshold
dynamic conditional correlations: An international analysis.
{\it Journal of Financial Econometrics} 11, 706-742.

\noindent Ling, S. (1999). On the probabilistic properties of a
double threshold ARMA conditional heteroskedastic model. {\it
Journal of Applied Probability} {\rm 36}, 688-705.

\noindent Ling, S. \& M. McAleer (2003). Asymptotic theory for a
vector ARMA-GARCH model. {\it Econometric Theory} {\rm 19}, 280-310.

\noindent L\"utkepohl, H. (1996). {\it Handbook of Matrices}. Wiley.

\noindent
Newey, W.K. (1997).
Convergence rates and asymptotic normality
for series estimators. {\it Journal of Econometrics} 79, 147-168.

\noindent Otranto, E. \& L. Bauwens (2015). Modeling the dependence
of conditional correlations on volatility. {\it Journal of Business \& Economic Statistics}. Available online.

\noindent Pelletier, D. (2006). Regime Switching for Dynamic
Correlations. {\it Journal of Econometrics} 131(1-2), 445-473.

\noindent Serre, D. (2010). {\it Matrices: Theory and Applications}.
GTM 216, Springer.

\noindent Tse, Y.K. \& A.K.C. Tsui (2002). A multivariate GARCH
model with time-varying correlations. \textit{Journal of Business
and Economic Statistics} {\rm 20}, 351-362.

\noindent Tweedie, R.L. (1988). Invariant measure for Markov chains
with no irreductibility assumptions. {\it Journal of Applied
Probability} {\rm 25}A, 275-285.

\pagebreak


\pagebreak

%
%
%
\end{document}